# Clinical Validation and Prospective Deployment of an Automated Deep Learning-Based Coronary Segmentation and Cardiac Toxicity Risk Prediction System

Christian V. Guthier, PhD[1,2],*; Christopher E Kehayias, PhD[1]; Cosmin Ciausu, MSc[1]; Jordan O. Gasho, BS[3]; John He, BA[1]; Maria Oorloff, BS[3]; Samuel C. Zhang, MD[3]; Danielle S. Bitterman, MD[1,2]; Jeremy S. Bredfeldt, PhD[1]; Kelly Fitzgerald, MD, PhD[1;] Benjamin H. Kann, MD[1,2]; David E. Kozono, MD, PhD[1]; Jennifer Steers, PhD[3]; Marion Tonneau, MD[2]; Anju Nohria, MD[4]; Hugo J.W.L. Aerts, Ph.D[1,2,5]; Katelyn M. Atkins, MD, PhD[3,**]; and Raymond H. Mak, MD[1,2,**]

1. Department of Radiation Oncology, Brigham and Women's Hospital/Dana-Farber Cancer Institute and Harvard Medical School, Boston, MA
2. Artificial Intelligence in Medicine (AIM) Program at Harvard-MGB, Boston, MA
3. Department of Radiation Oncology, Cedars-Sinai Medical Center, Los Angeles, CA
4. Division of Cardiology, Brigham and Women's Hospital/Dana-Farber Cancer Institute and Harvard Medical School, Boston, MA
5. Radiology and Nuclear Medicine, CARIM & GROW, Maastricht University, Maastricht, Netherlands

*Corresponding Author

** Co-Senior Author

Funding:

RHM: NIH-USA U01CA209414, JH: NIH-USA 5U01CA209414, KMA: Garber Award for Cancer Research, DSB: The American Cancer Society and American Society for Radiation Oncology, ASTRO-CSDG-24-1244514-01-CTPS Grant DOI #: https://doi.org/10.53354/ACS.ASTRO-CSDG-24-1244514-01-CTPS.pc.gr.222210; U54CA274516-01A1, HA: NIH-USA U24CA194354, NIH-USA U01CA190234, NIH-USA U01CA209414, and NIH-USA R35CA22052) ; BHK: NIH-USA U54 CA274516, NIH-USA P50 CA165962, NIH-USA R01 DE034780, St. Baldrick's Foundation Research Grant

Target: JAMA Oncology

**ABSTRACT**

**Importance:** Coronary artery radiation exposure is a strong predictor of major adverse cardiac events (MACE) and all-cause mortality (ACM) after thoracic radiotherapy, yet manual contouring is labor-intensive and therefore rarely performed in routine care.

**Objective:** To develop, externally validate, and clinically deploy an artificial intelligence (AI) auto-segmentation algorithm for cardiac sub-structures and perform prospective real-time surveillance of cardiac dose exposure.

**Design, Setting, and Participants:** Retrospective/prospective study to clinically validate AI auto-segmentation. A 3D UNet was trained on 560 thoracic planning CT scans from a single institution (2003-2014) and validated internally (n=70). External validation was performed in 283 patients with lung/breast cancer treated at an independent institution (2005-2020). Clinical implementation comprised (1) retrospective analysis of 3,399 lung cancer patients treated in 2014-2022 and (2) prospective real-time surveillance of 1,386 consecutive patients in 2023.

**Exposures:** AI–generated contours of the whole heart, chambers, and coronary arteries.

**Main Outcomes and Measures:** Geometric accuracy (Dice coefficient; average symmetric surface distance [ASSD]); concordance of dose-volume parameters; association of AI-derived left anterior descending coronary artery volume receiving ≥ 15 Gy (LAD V15) with MACE and ACM; temporal dose trends; and the proportion of patients exceeding prespecified high-risk thresholds (LAD or left circumflex [LCX] V15 ≥10% or mean heart/coronary dose ≥10 Gy).

**Results:** Median (inter-quartile range) Dice/ASSD were 0.95 (0.94-0.96)/1.1 mm for the heart and 0.87 (0.82-0.90)/1.9 mm for the LAD; the median absolute difference between AI and manual LAD V15 was 1% (Spearman ρ = 0.96). AI-derived LAD V15 remained independently associated with MACE (sub distribution hazard ratio [HR], 1.03/%; 95% CI, 1.01-1.05) and ACM (adjusted HR, 1.02; 95% CI, 1.00-1.03), internally and externally. Retrospective deployment showed a 32% relative decline in median LAD V15 from 2014 to 2022 (12% to 8%) and identified high- risk coronary/heart doses in 1,086 of 3,399 patients (32%). Prospective surveillance flagged 264 of 1,386 contemporary patients (19%) for potential cardiology referral.

**Conclusions and Relevance:** A validated AI-learning system accurately segments cardiac sub-structures substructures on routine CT scans, reproduces dose–outcome relationships, and enables large-scale surveillance and point-of-care alerts for high-risk patients. Automated cardiac dose monitoring could facilitate broader adoption of coronary-sparing radiotherapy and enhanced cardiovascular follow-up.

---

**Key Points**

- **Question:** Can deep learning automate cardiac sub-structuresubstructure contouring with sufficient accuracy for clinical decision support in radiotherapy?
- **Findings** The artificial intelligence algorithm achieved high geometric accuracy (median Dice 0.95 for the heart; 0.87 for the LAD), reproduced known associations between LAD dose and both MACE and mortality, and, when deployed, identified high-risk coronary doses in ~1 in 5 contemporary patients.
- **Meaning**: Automated, real-time cardiac dose surveillance is feasible and may accelerate adoption of coronary-sparing radiotherapy techniques and targeted cardio-oncology follow-up.

## INTRODUCTION

The rapid development of artificial intelligence (AI) techniques including deep learning-based computer vision algorithms, has the potential to transform cancer treatment delivery, particularly for radiotherapy (RT). AI algorithms can automate RT planning, including tasks such as manual segmentation of tumor and critical organs, as well as RT plan creation.[1] For instance, we have previously shown that lung tumor and cardiac auto-segmentation algorithms can perform at a level that matches inter-clinician variability[2,3] and AI algorithms can support quality assurance[4]. Moreover, we have demonstrated in clinical validation studies that human performance improves with access to these AI algorithms, leading to increased task speed and improved quality through reduction in unwarranted variation.[2] While many of these automation AI algorithms are entering the clinic as FDA-approved software as a medical device,[5] or as homegrown implementations,[3,6] the clinical validation of these algorithms is infrequently disclosed with scarce details provided.

Emerging evidence suggests that radiation dose exposure to the heart, and specifically the left coronary arteries, increases the risk of major adverse cardiac events (MACE), including for lung cancer,[7,8]_breast cancer,[9,10], and esophageal cancer.[11,12] While reduction of coronary dose exposure may mitigate these risks, there has not been widespread adoption of coronary or cardiac sub-structure sparing RT approaches in national guidelines.[13] The reasons for this lack of adoption are likely multi-factorial, but may include deficits in training, standardization and dissemination of best practices.[14] Furthermore, large scale research to accurately quantify the radiation dosimetric thresholds for coronary injury has been limited to atlas-based or modeling-based dose estimates because cardiac substructure segmentation is both technically challenging and time consuming.[15]

While there have been several publications including coronary artery auto-segmentation algorithms to address these gaps in clinical practice,[16–21] these studies are largely limited by small training sets, incomplete clinical validation, and absence of clinical implementation.

In this study, we report a comprehensive approach for end-to-end validation and clinical implementation of an AI-based auto-segmentation algorithm for coronary substructures. This approach incorporates standard volumetric validation with radiation dosimetric assessments and clinical outcome (cardiac events, survival) validation in independent and external datasets. Further, to comprehensively assess the potential clinical impact, we developed an automated AI deployment platform to test the algorithm in two real-world clinical and research settings: 1) large-scale retrospective analysis of temporal trends of cardiac sub-structure dose exposure at the population level; and 2) prospective clinical deployment to automatically identify patients with high cardiac sub-structure radiation dose exposure and notify physicians for decision support and quality improvement.

## METHODS

### *Clinical Datasets*

Several large datasets were utilized in this study (**Figure 1A**) and patient characteristics are reported in **Table 1**. First, a dataset of 748 patients with locally advanced non-small cell lung cancer (NSCLC) treated with thoracic RT from 2003-2014 at Brigham and Women's Hospital/Dana-Farber Cancer Institute (BWH/DFCI) was used for AI training. Pre-treatment clinical characteristics, cardiac sub-structure dosimetric data, and outcomes including survival and cardiac toxicity have been previously reported.[7,22,23] Briefly, whole heart and cardiac sub-structures including atria, ventricles, and coronary vessels (left main, left anterior descending [LAD], left circumflex [LCX], right, and posterior descending coronary arteries) were manually segmented on RT planning CT scans based on atlas guidelines [24,25]. The RT planning CT scans include both non-contrast and contrast-enhanced studies as well as free-breathing and 4-dimensional CT images. Of these 748 patients, 48 were excluded from algorithm development due to incomplete data (e.g., missing dose or structures), and the remaining 700 were split into training (n=560), validation (n=70), and testing (n=70) groups.

The external validation set (n=283) included patients with locally advanced NSCLC (n=102) or breast cancer (n=181) treated with thoracic RT between 2005 and 2020 at Cedars-Sinai Medical Center, with paired clinical, dosimetric and outcomes data. The LAD was retrospectively segmented, as previously described,[26] per atlas guidelines.[7,24,25]

For the large-scale retrospective and prospective deployments (**Figure 1B-1C)**, internal datasets were generated from sequential patients within the BWH/DFCI system where the RT planning CT included the heart. The retrospective dataset (n=3,399) included sequential patients with NSCLC treated with thoracic RT between January 2014 and October 2022. The prospective dataset (n=1,386) included sequential patients with any malignancy treated with RT and with the heart imaged in the planning CT scan between January-December 2023.

All datasets were analyzed under a data use agreement and institutional review board protocols from Dana-Farber/Harvard Cancer Center (DF/HCC) and Cedars-Sinai Medical Center (DF/HCC 11-286/20-328; CSMC STUDY00000759).

***Cardiac Sub-Structure Auto-Segmentation Algorithm Training***

The deep learning-based **R**adiation **O**ncology **Card**iac Sub-Structure **S**egmentation (RO-CardS) algorithm was trained using a 3D U-Net with ResNet encoders and cross connections to localize the heart and segment its substructures, which included the four chambers and left main, LAD, LCX, right and posterior descending coronary arteries (see **Supplemental Methods** for additional details)[27].

***Geometric and Dosimetric Validation***

Algorithm performance was assessed and validated by comparing the AI-generated segmentation versus the manual ground truth segmentation using geometric analyses with the Dice coefficient (volumetric overlap) and average symmetric surface distance (ASSD –average distance between the surfaces of two segmentations). These two measures are complementary, since even a small deviation for small structures such as the coronaries could lead to significant changes in volumetric overlap, but does not necessarily reflect poor quality segmentation in the clinical context.[28]. Dosimetric validation was performed by evaluating mean, maximum, and volume (V) receiving 15 Gy (V15Gy), a dosimetric threshold previously identified as associated with increased risk of MACE[7]. The Kolmogorov-Smirnov test was also calculated to evaluate the similarity of dose-volume histograms (DVH) generated by manual versus AI segmentations.

***Clinical Outcomes Validation***

Clinical outcomes validation was performed by inputting the cardiac substructure dose calculated from the AI-generated cardiac sub-structure segmentations, including the AI-predicted

LAD V15, into our previously published risk prediction models in both the internal and external validation datasets [7,22]. Clinical endpoints included all-cause mortality (ACM) and major adverse cardiovascular events (MACE)[29]—a validated cardiac endpoint defined as a composite of heart failure hospitalization or urgent visit, unstable angina, myocardial infarction, coronary revascularization, and cardiac death. Methods for collection of clinical risk factors and endpoints are described in the **Supplemental Methods**. MACE cumulative incidence estimates were adjusted with non-cardiac death as a competing risk and compared using a 2-sided Gray p-value. Kaplan-Meier estimates of all-cause mortality were compared using the log-rank test. Cox and Fine and Gray regressions were performed (adjusted for non-cardiac death as a competing risk. A 2-sided p-value ≤0.05 was considered statistically significant. All analyses were performed using Stata, version 17.0 (StataCorp LLC).

***Clinical Application #1: Large-Scale Retrospective Assessment of Cardiac Sub-Structure Dose Trends Over Time***

To test a clinical application of monitoring cardiac sub-structure dose exposure trends at the population level for quality improvement, the ROCardS algorithm was deployed on a large retrospective cohort (n=3,399; **Figure 1B**) to auto-segment the LAD, LCX, left ventricle (LV), and whole heart on RT planning CT scans. A software platform was developed to automatically calculate radiation dose-volume relationships for each AI-generated cardiac sub-structure with the RT plan that was clinically delivered, using methods previously described.[6] For patients receiving hypofractionated RT, including stereotactic body RT (SBRT), a biologically equivalent dose was calculated for all dose-volume estimates (equivalent dose in 2 Gy fractions [EQD2] with $\alpha/\beta=3$). AI-generated sub-structure dose exposure trends over time were analyzed at published, clinically-relevant dose thresholds (e.g. LAD, LCX, and LV V15; mean coronary dose, mean heart dose [MHD]) [7,22], and stratified by clinically-relevant factors including location of tumor (left versus right), and RT technique (intensity modulated RT [IMRT] versus 3D-conformal RT [3D-CRT]).

Mann-Kendall statistical tests for trend were calculated to assess changes in dose exposures over time.

### *Prospective Clinical Deployment to Screen for Patients with High-Risk Cardiac Dose Exposure*

Next, to prospectively test the system to automatically screen for patients with high cardiac sub-structure radiation dose exposure, the ROCardS system was clinically implemented for ambient cardiac dose exposure monitoring and physician decision support. The system automatically monitors all new RT plans for the presence of a manually segmented heart daily prior to treatment (**Figure 1C**). If heart segmentation is present, the system automatically downloads CT images, RT plan, and dose data, deploys ROCardS auto-segmentation, and calculates sub-structure dose. The system then auto-generates a report summarizing clinically relevant cardiac sub-structure radiation dosimetric data, which is then sent to a departmental quality improvement (QI) team for review (**Supplemental Figure 1** for a sample report). Further details regarding this AI deployment system were reported previously.[6] The system was prospectively deployed from January-December 2023 in the BWH/DFCI clinic under an IRB-approved QI protocol (DF/HCC 20-2328).

## RESULTS

***Geometric and Dosimetric Validation of ROCardS Auto-Segmentations in the Internal Dataset***

Total processing/segmentation time of the ROCardS algorithms was <1 minute (**Supplemental Results**). Geometric validation results (Dice and ASSD) are shown in **Supplemental Table 1**. The performance, as measured by volumetric overlap (Dice), decreased with decreasing cardiac sub-structure size. Heart auto-segmentation had the best performance (Dice 0.95 [IQR=0.01]), while the worst performance was observed with left main coronary auto-segmentation (Dice 0.40 [IQR=0.29]). In contrast, the ASSD was between 1-2 mm for all structures, showing sufficient agreement for this radiation oncology use case with performance independent of sub-structure size. **Supplemental Figure 2** provides representative examples of cardiac sub-structure auto-segmentations.

The radiation dosimetric data from AI-generated versus manual cardiac sub-structure segmentations were highly correlated (**Supplemental Table 1**). For example, the difference in mean dose calculated from AI versus manual segmentations was -0.03 (IQR=0.8) Gy for the heart, -0.04 (IQR=0.30) Gy for the LV, and -0.01 (IQR=0.14).2 Gy for the LAD. For the clinically-relevant LAD V15, the median difference was -1% (IQR=5%). There was no significant difference in the overall distribution of the dose-volume histograms for nearly all cardiac substructures from AI versus manual segmentations (Kolmogorov-Smirnov median p-value >0.05), except for the left main coronary (median p=0.029, but with 45% of cases showing no significant difference).

*Internal Validation of ROCardS for Clinical Outcomes Prediction*

AI-generated LAD V15 Gy dose estimates replicated the previously reported associations with both MACE and ACM that were observed using manually-generated LAD segmentations[7]. In the BWH/DFCI internal validation cohort (n=70), adjusting for age, sex, pre-existing hypertension (HTN), and coronary heart disease (CHD), there was an increased risk of MACE with AI-generated LAD V15 Gy (sub-distribution hazard ratio [sHR] 1.03 per %; 95% CI, 1.01-1.05; p=.014) (**Figure 2; Table 2; Supplemental Table 2**). In the same model, LAD V15 Gy from manual research segmentations was similarly associated with MACE (sHR 1.05 per %; 95% CI, 1.03-1.07; p<0.001; **Supplemental Table 3**). Similarly, after adjusting for lung cancer and cardiac prognostic factors, AI-generated LAD V15 Gy was associated with an increased risk of ACM (adjusted HR [aHR] 1.02 per %, 95% CI, 1.00-1.03; p=0.046) (**Supplemental Table 4**), closely replicating the association observed with the ACM model using manually-segmented LAD V15 Gy (aHR 1.03 per %; 95% CI, 1.01-1.05; p=.006) (**Supplemental Table 5**).

***External Validation of Geometric, Dosimetric, and Clinical Outcomes***

In the external dataset (n=283), AI-generated cardiac sub-structures had high geometric overlap with manual segmentations with ASSD values ranging from 1.3-2.7 mm and the radiation dosimetry for AI versus manual segmentations was also highly correlated (**Supplemental Table 6)**. Clinical characteristics for the external dataset are reported in the **Supplemental Results** and **Table 1,** and the 2-year cumulative incidence of MACE was 5.1% (95% CI, 3.0-8.0%). AI-generated sub-structures in this external dataset replicated the association of LAD V15 with key clinical outcomes. Accounting for pre-existing CHD, there was an increased risk of MACE with AI-predicted LAD V15 Gy (SHR 1.02 per %; 95% CI [1.00-1.03]; p=.045; **Table 2; Figure 2; Supplemental Table 7**). Accounting for age, sex, performance status, pre-existing HTN, and CHD, there was an increased risk of ACM with AI-generated LAD V15 Gy (aHR 1.02 per %; 95% CI 1.02 [1.01-1.03]; p=.001; **Table 2; Supplemental Table 8, Supplemental Figure 3**).

***Clinical Application #1: Large-Scale, Real World Temporal Trends in Cardiac Sub-structure Dose for Lung Cancer RT***

Large-scale retrospective deployment of ROCardS on an independent dataset of 3,399 lung cancer patients was performed to calculate cardiac substructure dosimetry trends (**Table 1**). There was a statistically significant decrease in all cardiac sub-structure doses from 2014 to 2022 (Mann-Kendall trend test, p<0.01 for LAD, LCX and MHD; p<0.05 for LV; **Figure 3A-C**). Notably, there was a significant decrease in LAD V15 that coincided with the publication of a prior study identifying the association of this dose variable with the risk of MACE and ACM in 2021.[7]

On sub-group analyses, there was significantly higher V15 exposure for left-sided tumors versus right sided tumors (p<0.01; **Figure 3D-F**). For both left- and right-sided RT plans, LAD, LCX and mean heart dose decreased significantly over time (p<0.01). By RT technique (**Figure 3G-I**), there was no significant difference in LAD V15 Gy with 3D-conformal technique over time (p>0.05 for trend), but a significant decrease in LAD V15 over time with IMRT and SBRT (p<0.01 for trend). As an example of population-based risk assessment, 32.0% (n=1086) of the patients within this cohort had high cardiac risk exposures (LAD V15 Gy ≥10%) (**Supplemental Table 9)**. Notably, only 2.3% (80/3398) of patients in this lung cancer cohort had any coronary sub-structure segmented as part of standard clinical care, illustrating substantial opportunity for improvement and leading to departmental implementation of standardized LAD/LCX constraints.

***Clinical Application #2: Prospective Clinical Deployment for Ambient Surveillance of Patients with High-Risk Cardiac and Coronary Dose Exposure***

The ROCardS system was prospectively deployed in the clinic for 12 months and analyzed 1,950 patients with various cancer diagnoses (**Table 1** for patient characteristics). The system automatically identified 317 patients (16.3%) with potentially high-risk exposures (LCX or LAD V15 Gy ≥10%, MHD ≥10 Gy, or mean coronary dose ≥10 Gy; **Figure 4A**). Among disease sub-groups, most esophageal cancers (88.8%), 34.1% of thoracic malignancies (mainly lung cancer), and a minority of breast cancer cases (0.7%) had high risk cardiac exposures, which is consistent with previously reported studies. While some lung cancer patients had high LAD and/or LCX exposures, esophageal cancers patients routinely had high LCX exposures (**Supplemental Figure 4**). Unexpectedly, a sizeable proportion of patients receiving spine radiation therapy also had high risk exposures (29.7%)., leading directly to a quality improvement initiative to standardize cardiac constraints on this group of patients. Examining the correlation between LAD V15 and MHD by disease sub-groups (**Figure 4B**), patients with thoracic, breast and spine treatment sites most commonly had discordant low MHD (<10 Gy) with high LAD exposures (V15 $\geq$10%), highlighting disease sites where MHD might not appropriately capture high risk exposures.

## DISCUSSION

Coronary artery radiation dose is an independent predictor of MACE and ACM following thoracic RT for LA-NSCLC. However, cardiac sub-structure segmentation is not routinely performed, in part due to technical complexity. We developed and validated a deep learning-based automated cardiac sub-structure contouring tool to address this gap. The key novelties of our study include: (1) AI training on a large, clinically-validated coronary artery segmentation dataset (n=748), (2) performing comprehensive geometric, dosimetric, *and* clinical validation of the AI-generated segmentations to predict a validated cardiac toxicity endpoint (MACE) and ACM in an independent dataset, as well as (3) large-scale retrospective and prospective clinical deployment to demonstrate practical, impactful use cases. Additionally, the algorithm was trained and externally validated on real-world CT datasets (including a mixture of contrast and non-contrast scans, and static CT and respiratory 4D-CT image sets, suggesting robust performance**.**

Many cardiac toxicity studies for lung cancer have focused on whole heart dose exposures[22,30,31] or been limited to mapping regions of interest such as the “heart base”, using physical dose-volume-based mapping and mainly associating those outcomes with survival, which is relatively non-specific[32,33]. While the heart base studies have been useful to help identify overlapping ‘top’ regions of the heart that are associated with composite cardiac events and overall survival, the region has been variably defined and does not precisely correlate with an anatomic definition (and typically includes the coronary origins, sinoatrial node, and atria), thereby precluding further insights into the pathophysiological mechanisms leading to specific cardiac toxicity sub-types, and limits translation and risk mitigation strategies. Work by us and others have shown that mapping radiation dose exposure to **anatomic** and physiologically-relevant cardiac substructures, in combination with assessment of baseline risk factors or pre-existing cardiac disease, results in identification of biologically-plausible and clinically actionable dose exposure thresholds[7,23,34]. Thus, the development of reproducible and automated anatomic cardiac sub-structure segmentation has the potential to accelerate our understanding of radiation-related cardiac injury in a variety of disease sites.

Furthermore, we illustrated the potential clinical impact of these auto-segmentation algorithms through two clinical uses cases. First, we implemented the algorithms across large real-world datasets of thousands of lung cancer patients to quantify population-level cardiac dose exposure and demonstrate trends over time, which can accelerate research, quality improvement and identify previously treated patients at highest risk for cardiac toxicity. We demonstrated that mean heart dose did not significantly change over time from 2014-2022, but LAD V15 decreased significantly in 2021-2022, with significantly more cases achieving a V15<10% compared to prior years. This change in LAD dose corresponded to the publication of our prior work demonstrating an association between LAD V15 > 10% and MACE and ACM. The AI system was able to capture this practice change, illustrating the potential for AI automated segmentation to transform how we capture and track radiation dose exposure over time. In comparison, prior large-scale population studies on cardiac toxicity in cancer patients treated with radiation therapy have relied on dose estimations or generalized atlases,[15,35,36] which have been practice changing in some cases, but our automated approach has the potential to provide much more accurate dose estimates at scale. Furthermore, illustrating the actionability of population level analyses, the identification of outlier right sided cases receiving unexpectedly high LAD exposures led to a practice-based quality improvement initiative to standardize dose constraints to the LAD/LCX for all lung cases.

Second, we prospectively deployed the algorithm as an ambient, background clinical surveillance and physician decision support tool that automatically identifies patients receiving radiation therapy with high-risk cardiac and/or coronary exposure. We designed this pilot approach to simultaneously address two potential barriers to widespread adoption of sub-structure dosimetric constraints: 1) lack of knowledge and training on how to segment specific sub-structures in a standardized manner; and 2) lack of trust in AI systems by clinicians. By providing ambient screening and flagging high risk patients for further intervention, our deployment delivered value-added clinical insights that could foster clinician trust and support wider clinical deployment. Again, we identified actionable clinical insights from this prospective AI deployment, such as the establishment of cardiac dose constraints for spine RT cases given the unexpectedly high dose exposures observed in our prospective surveillance study.

Implementation science analyses reveal that high-value radiotherapy innovations take many years to reach patients with the median time from pivotal publication to 50% clinical uptake ranging from **8 to 17 years** [14]. Examples of this evidencetopracticeevidence-to-practice gap include deep-inspiration breath-hold and single-fraction palliative bone RT, which illustrate how equipment costs, training demands, and misaligned incentives impede adoption despite robust evidence. Coronary-sparing RT encounters the same obstacles: few clinicians routinely segment the LAD or LCX, segmentation techniques and training vary, and guideline updates trail emerging data. Our open source automated coronary-segmentation algorithm and real-time dose-surveillance platform removes the manual segmentation bottleneck, integrates seamlessly into existing workflows, and delivers actionable alerts to directly address key barriers to adoption (complexity, workflow fit, resource burden). In doing so, it has the potential to shorten the evidence-to-practice gap and make coronary dose constraints a standard component of thoracic radiotherapy planning.

## CONCLUSION

We developed deep learning automated cardiac sub-structure segmentation algorithms from a large real-world planning CT dataset—including contrast and non-contrast, non-gated CT scans. The algorithms produced high quality contours with high geometric overlap with manual contours and were validated with dosimetric, cardiac toxicity, and survival outcomes. We demonstrated that high quality auto-segmentation enables multiple real-world clinical and research applications and holds significant potential for near-term clinical translation to improve cardiac toxicity risk prediction.

## FIGURES

**Figure 1.** Study Design, including a) deep learning cardiac sub-structure auto-segmentation algorithm training and validation in internal and external data sets utilizing volumetric and dosimetric measures and clinical outcomes; b) real-world, large-scale clinical deployment of automated deep learning-based cardiac substructure segmentation and automated radiation dose exposure calculation on 3398 cancer patients to identify temporal trends in cardiac dodge exposure at population level and identify patients for cardiac risk mitigation; c) prospective deployment of the deep learning-based automated cardiac substructure dose exposure calculation system in the clinic as a quality assurance system to identify patients with high risk cardiac exposure for risk mitigation.

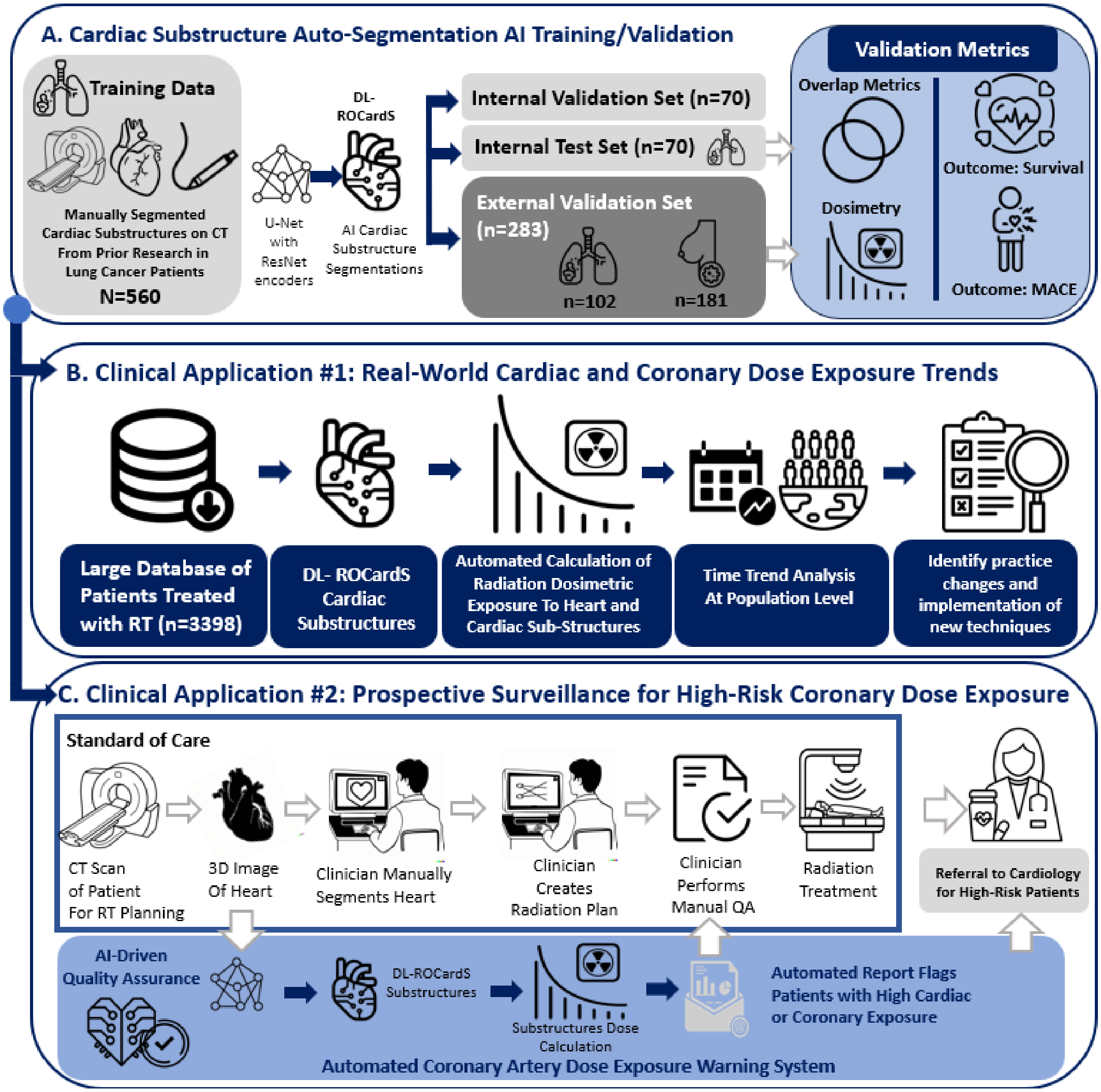

A. Cardiac Substructure Auto-Segmentation AI Training/Validation
Training Data
Manually Segmented Cardiac Substructures on CT From Prior Research in Lung Cancer Patients
N=560
U-Net with ResNet encoders
DL-ROCardS
AI Cardiac Substructure Segmentations
Internal Validation Set (n=70)
Internal Test Set (n=70)
External Validation Set (n=283)
n=102
n=181
Validation Metrics
Overlap Metrics
Dosimetry
Outcome: Survival
Outcome: MACE
B. Clinical Application #1: Real-World Cardiac and Coronary Dose Exposure Trends
Large Database of Patients Treated with RT (n=3398)
DL- ROCardS Cardiac Substructures
Automated Calculation of Radiation Dosimetric Exposure To Heart and Cardiac Sub-Structures
Time Trend Analysis At Population Level
Identify practice changes and implementation of new techniques
C. Clinical Application #2: Prospective Surveillance for High-Risk Coronary Dose Exposure
Standard of Care
CT Scan of Patient For RT Planning
3D Image Of Heart
Clinician Manually Segments Heart
Clinician Creates Radiation Plan
Clinician Performs Manual QA
Radiation Treatment
Referral to Cardiology for High-Risk Patients
AI-Driven Quality Assurance
DL-ROCardS Substructures
Substructures Dose Calculation
Automated Report Flags Patients with High Cardiac or Coronary Exposure
Automated Coronary Artery Dose Exposure Warning System

**Figure 2.** Cumulative incidence of MACE by LAD V15 Gy ≥10% vs <10% in (A) internal validation cohort (n=70) with AI-generated LAD V15 Gy (left) which replicates the results of manual LAD segmentations (right). (B) External validation of LAD V15 Gy AI-generated prediction in external validation cohort (n=283).

A. BWH internal validation cohort, AI-generated LAD V15 Gy (left), manual LAD V15 Gy (right)

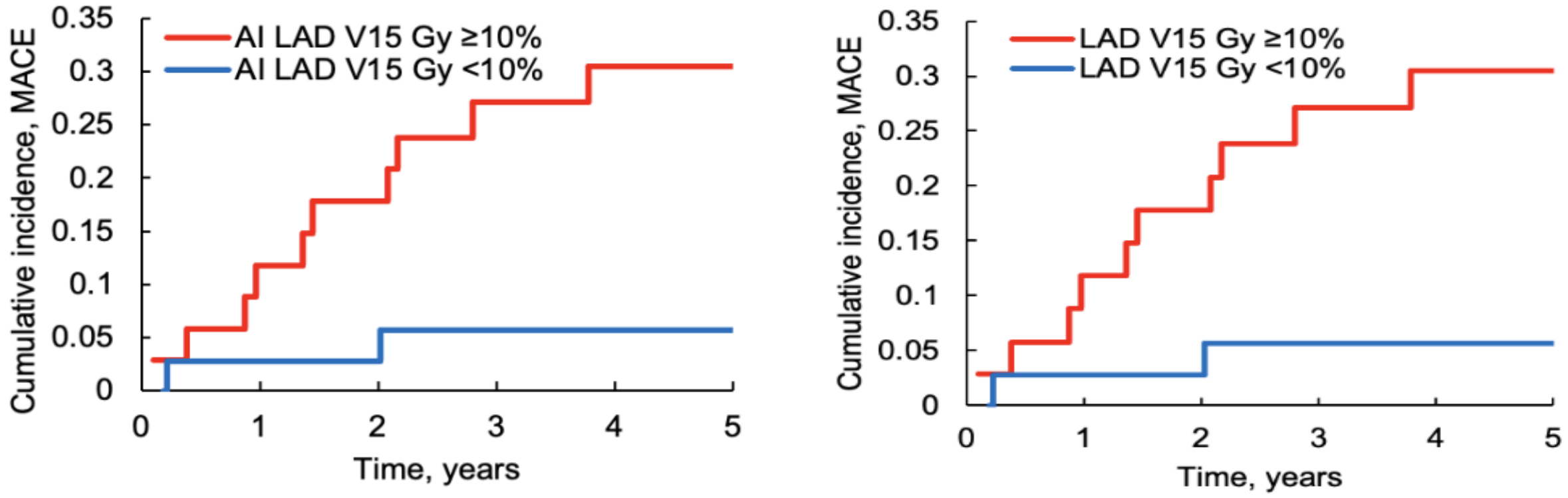

B. AI-predicted LAD V15 Gy in CSMC External Validation Cohort

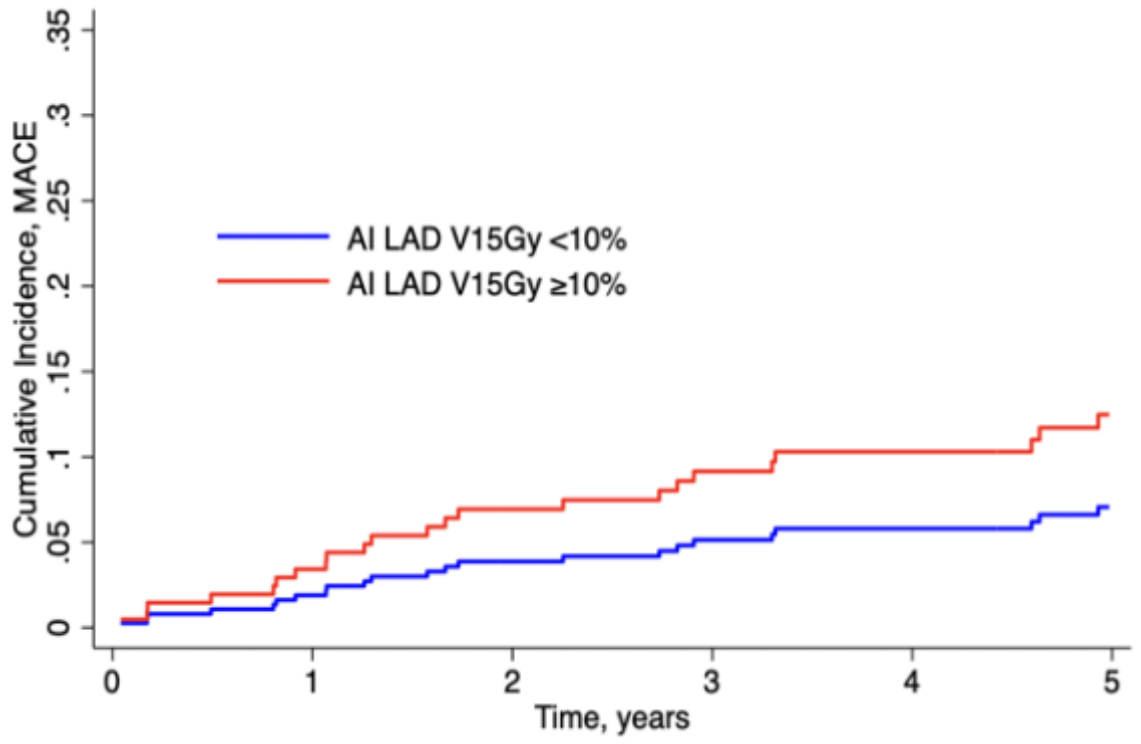

**Figure 3.** Temporal trends by year (2014-2022) of cardiac sub-structure radiation dose exposure in a large, real-world lung cancer dataset (n= 3399) stratified by tumor location and radiotherapy (RT) technique. (**A-C**) Temporal trends for V15 Gy for the left anterior descending (LAD) and left circumflex (LCX) coronary arteries, and mean heart dose (MHD); **(D-F)** trends by left vs right tumor location; and (**G-I**) trends by intensity modulated RT (IMRT) vs 3-dimensional RT (3D-CRT) vs. Stereotactic body radiation therapy (SBRT) technique.

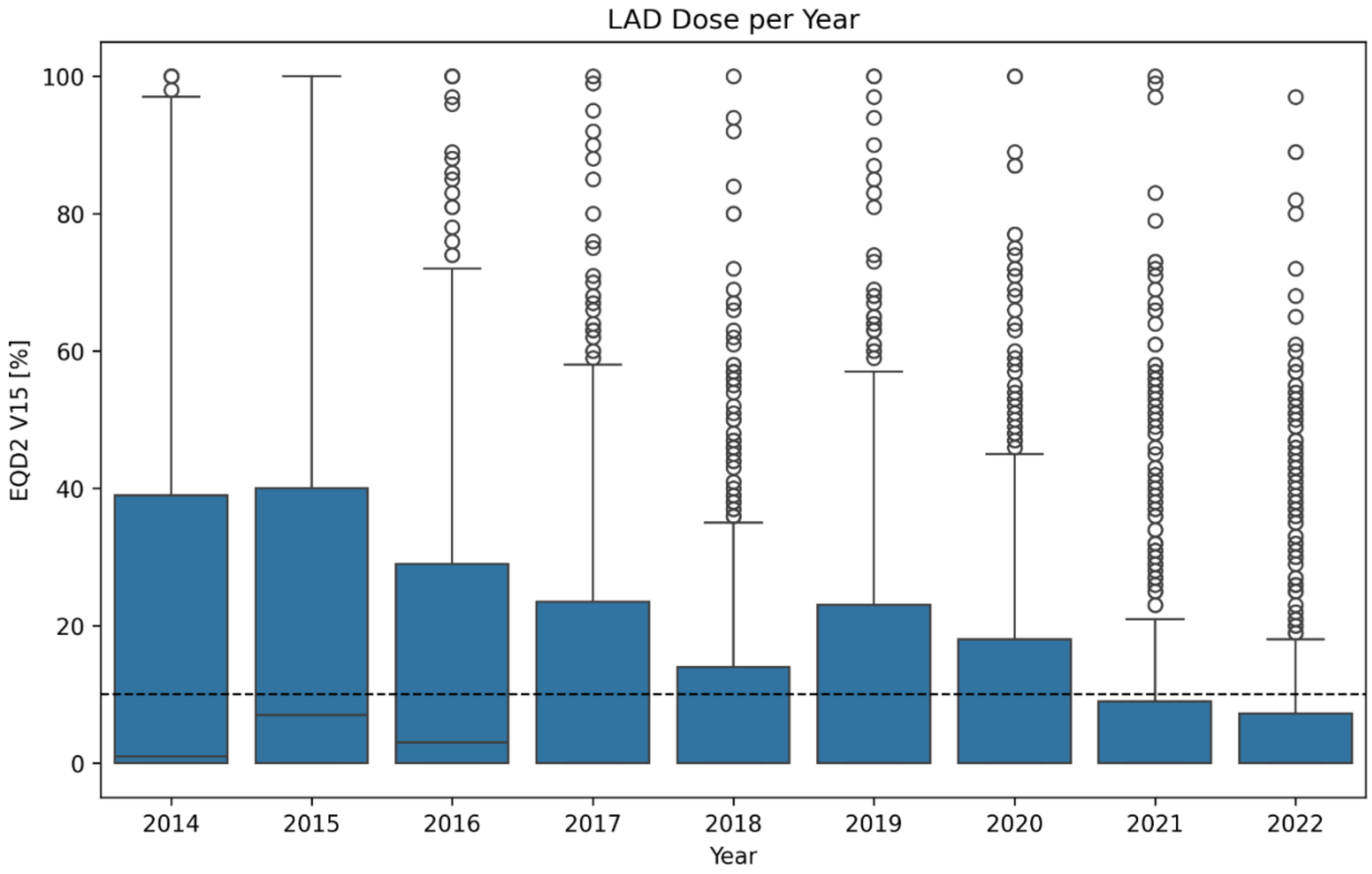

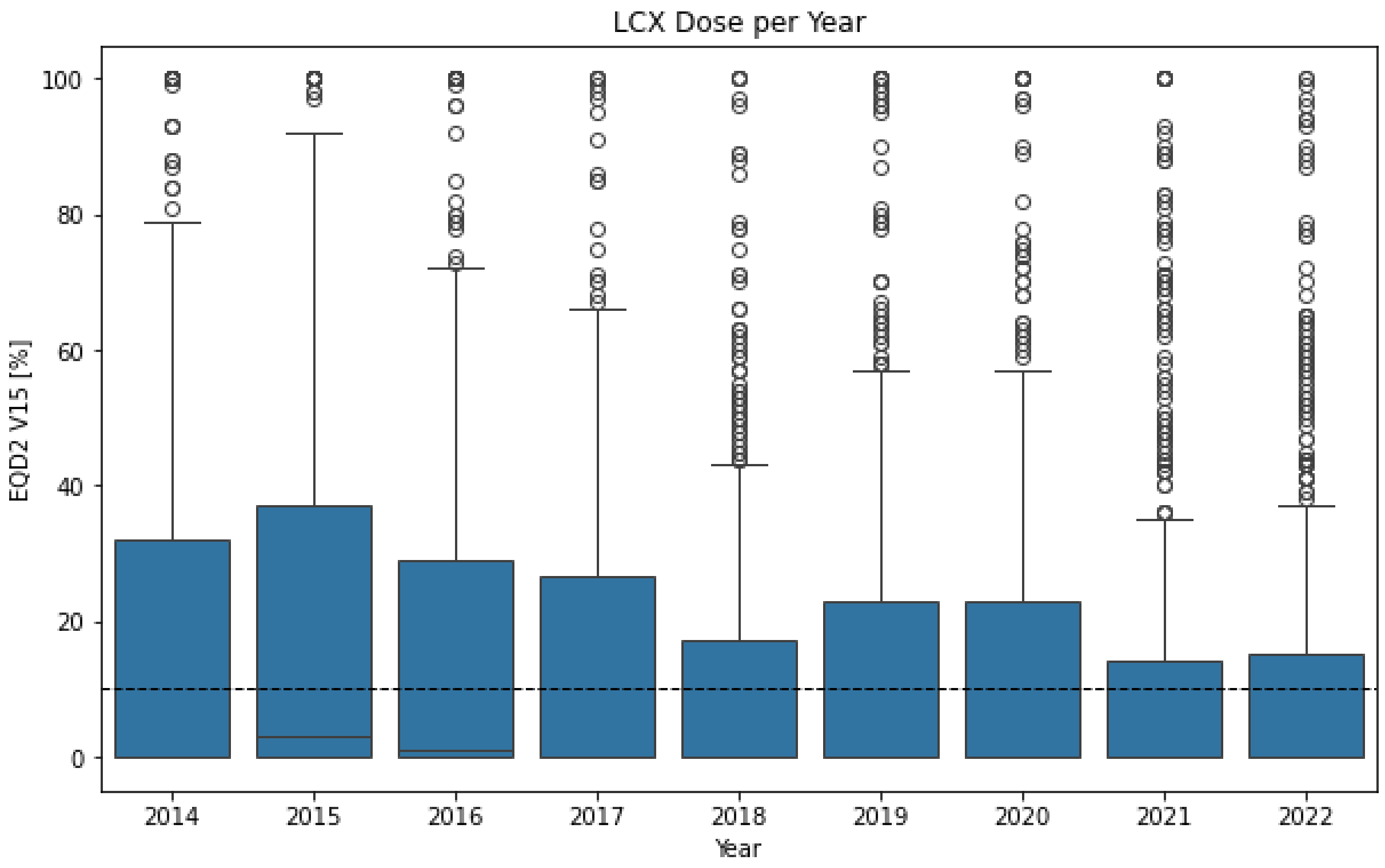

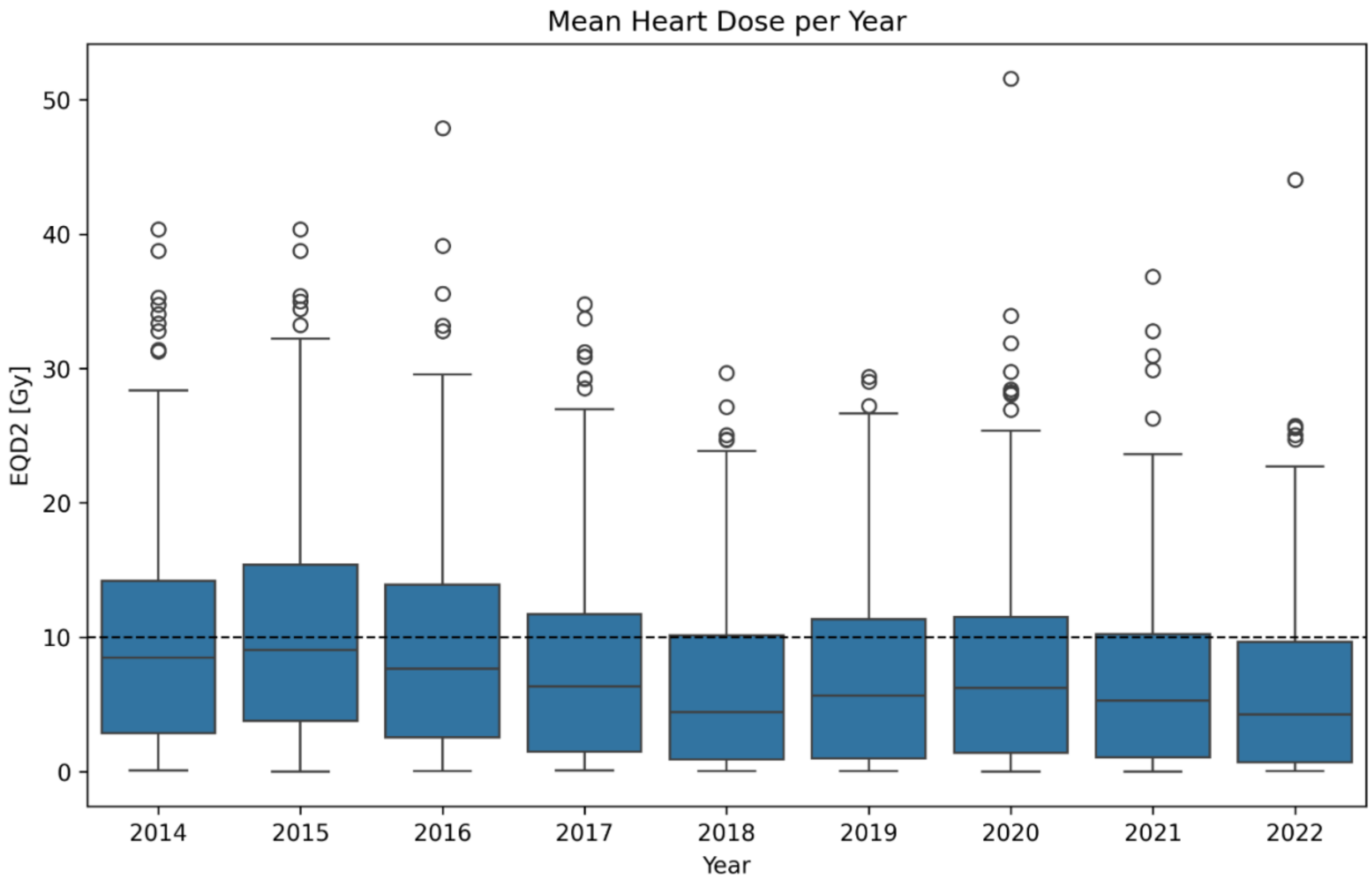

**D-F** (left vs right)

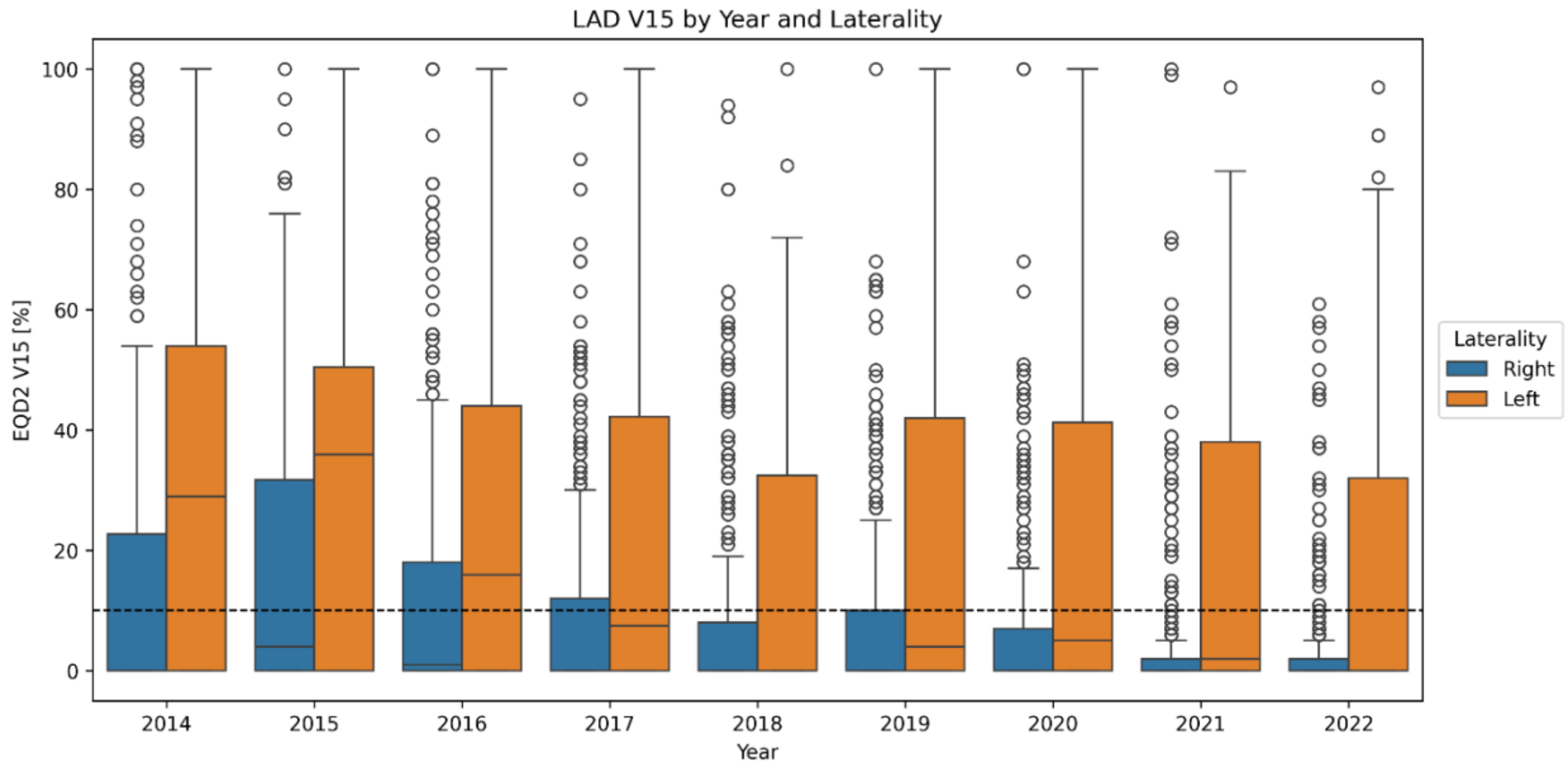

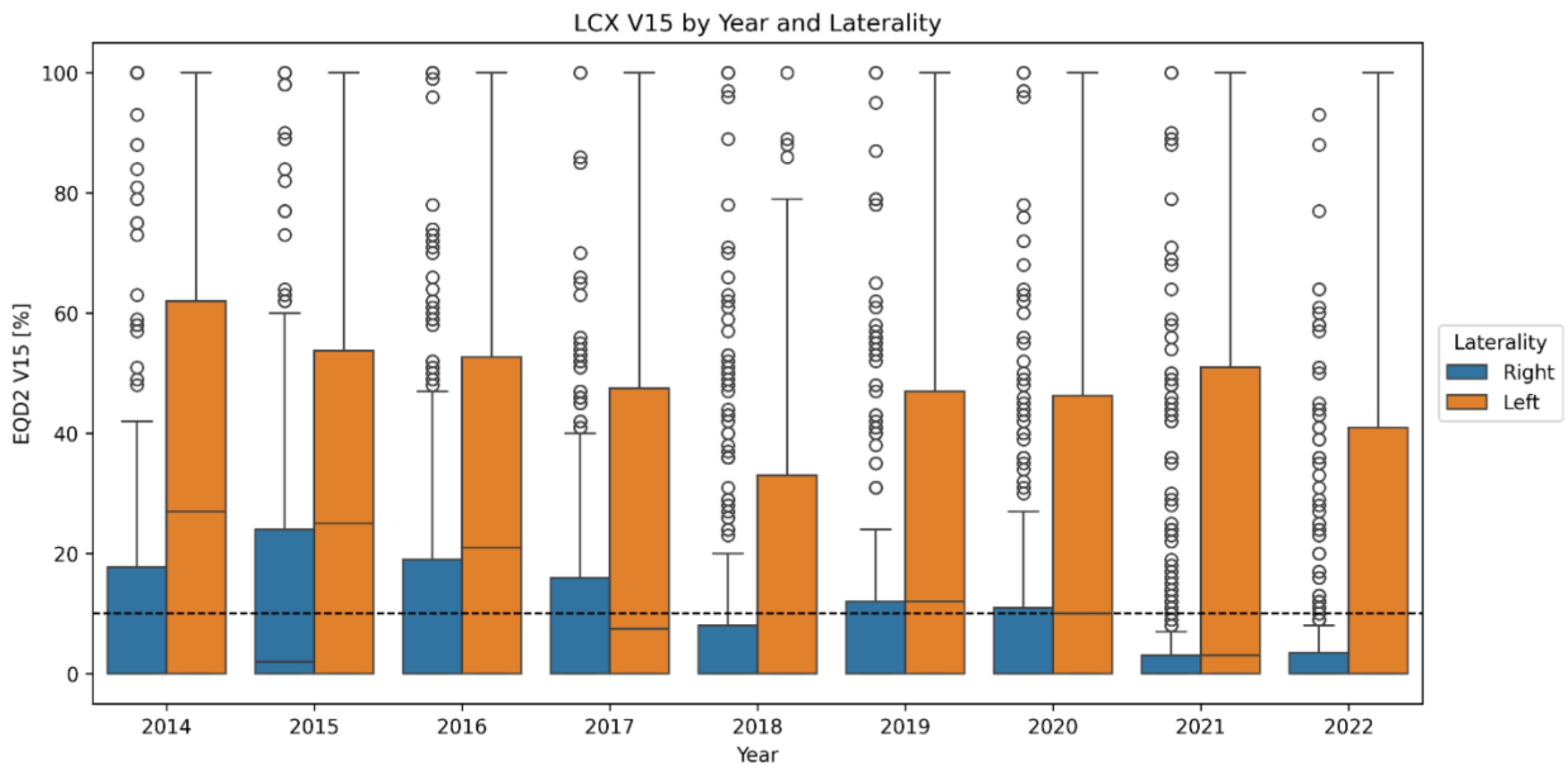

**G-I** (IMRT vs 3D-CRT vs SBRT)

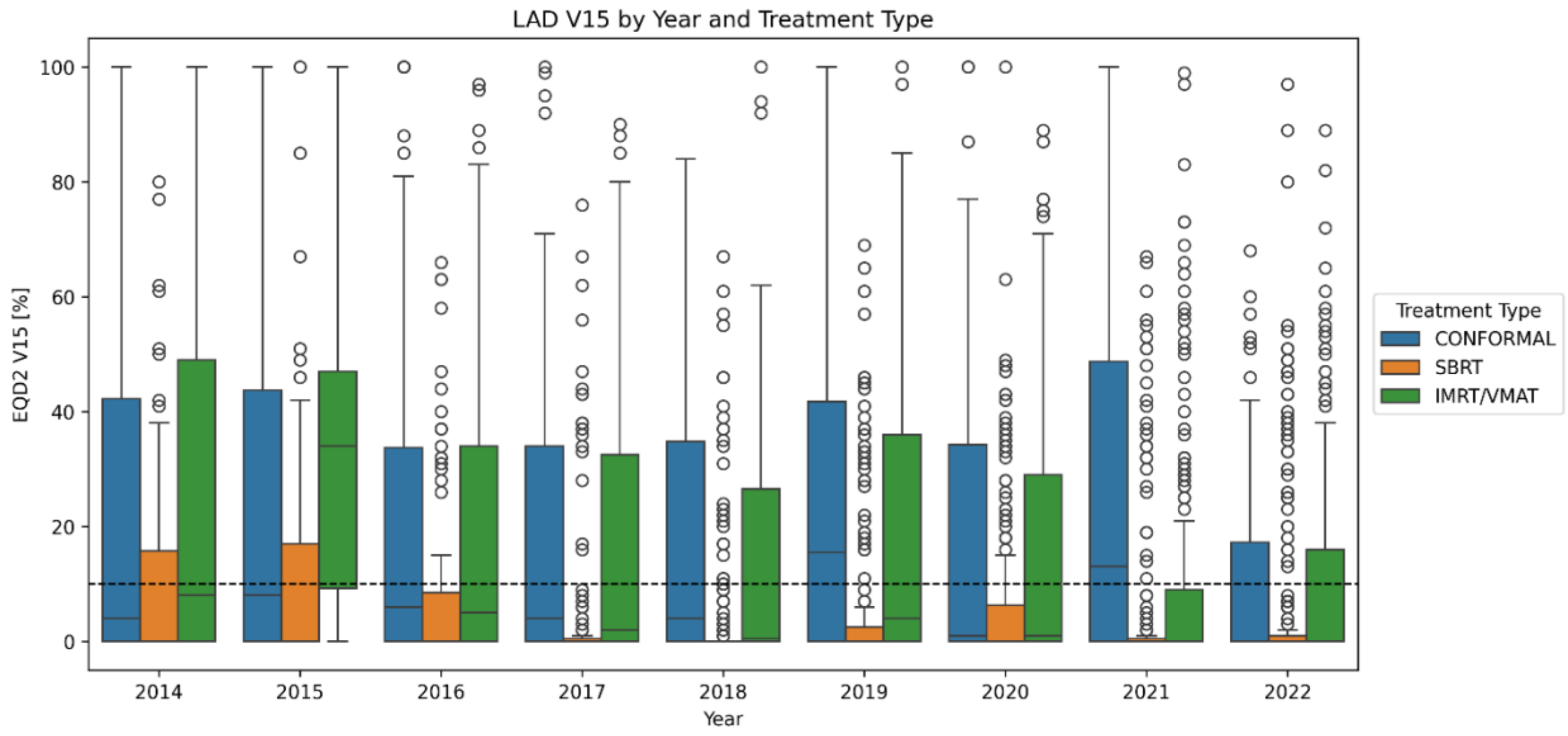

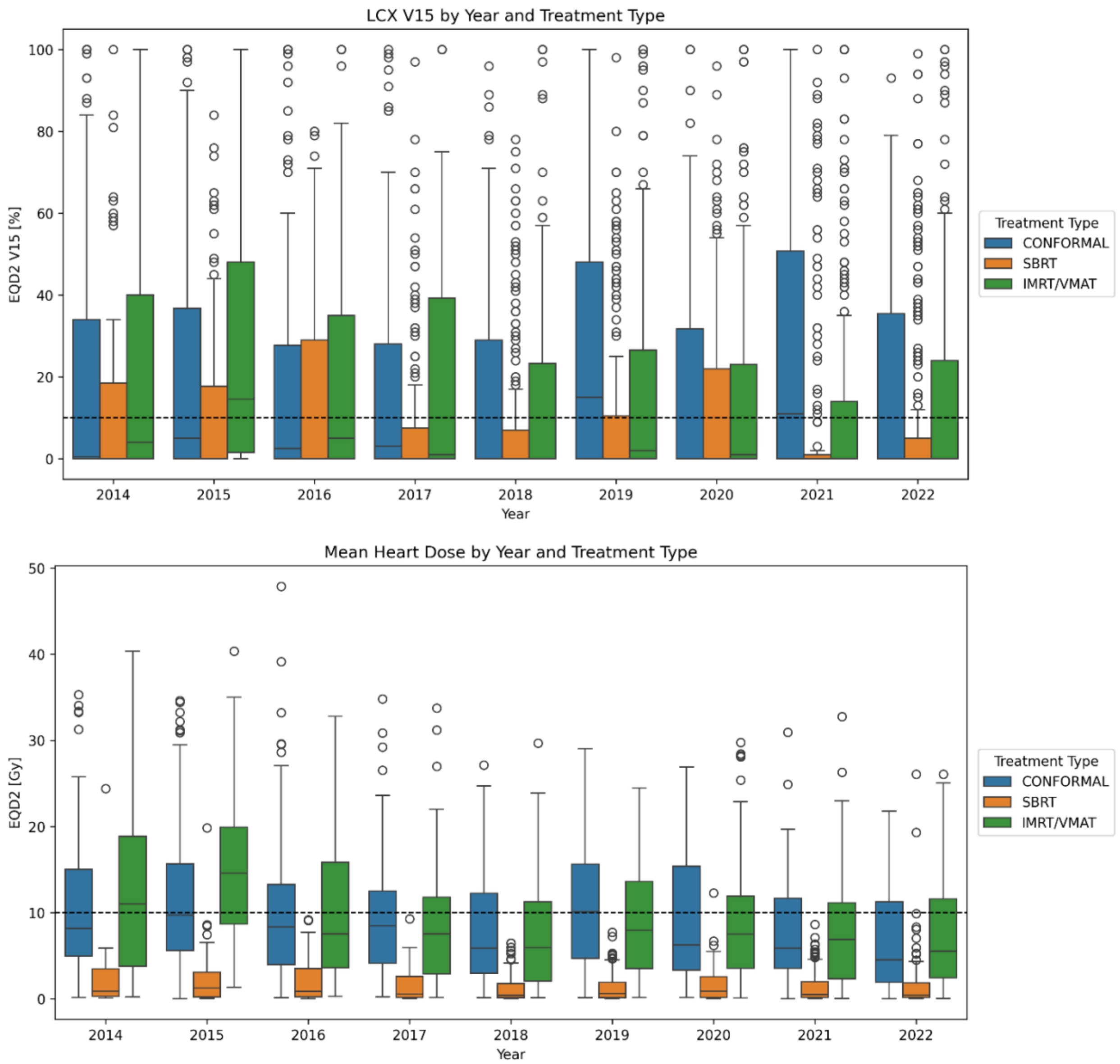

**Figure 4**. Identification of high-risk cardiac radiation exposure. Prospective clinical deployment of the deep learning-based automated cardiac sub-structure segmentation and dose exposure system (ROCardS) with ambient analysis of 1950 patients with cardiac radiation exposure, identifying 317 patients with high-risk dose exposures to either the LCX or LAD (V15 ≥10%) or mean heart or coronary dose (≥10 Gy). (A) High cardiac risk dose exposures by disease site.

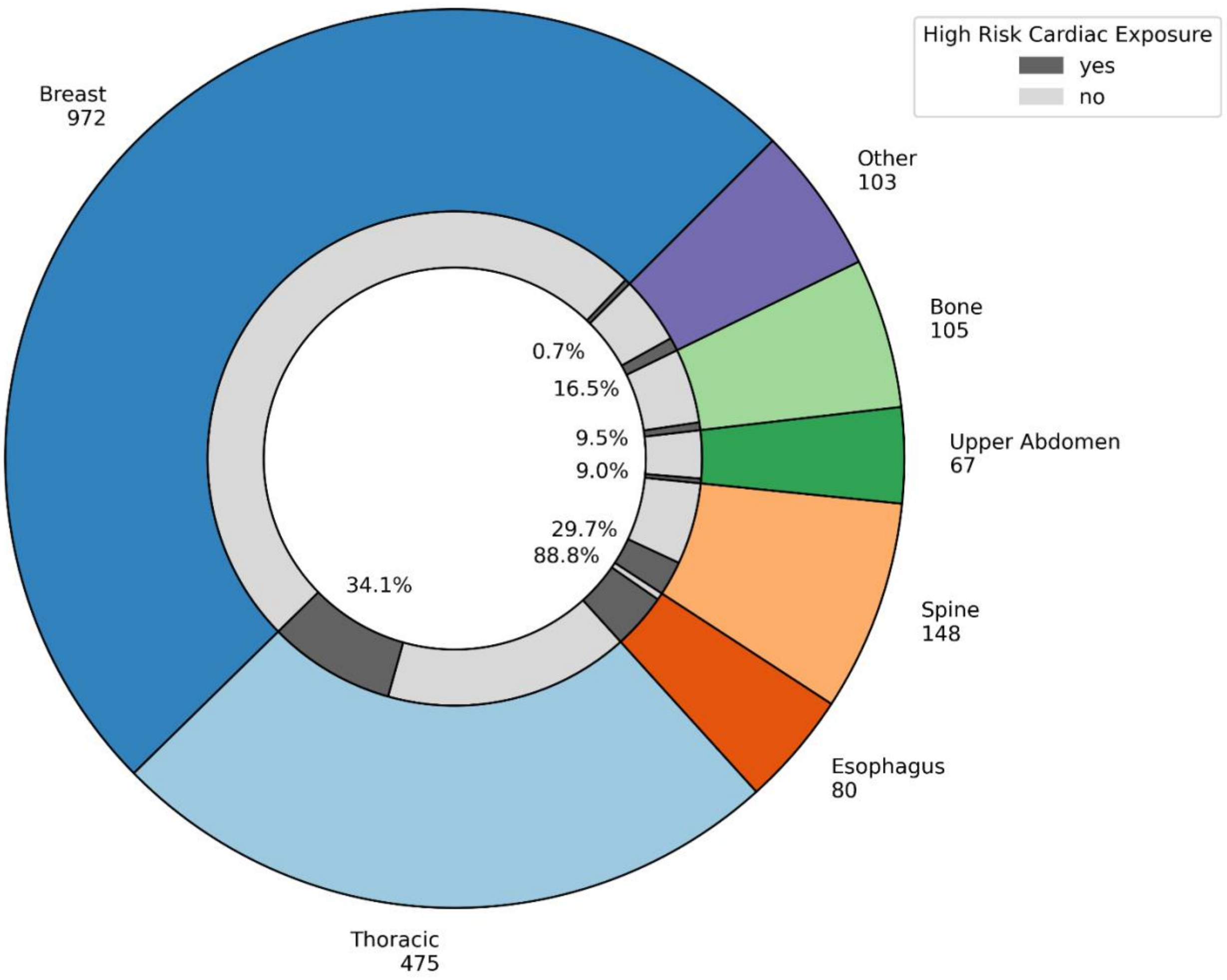

(B) scatter plot of MHD vs LAD V15 by disease site.

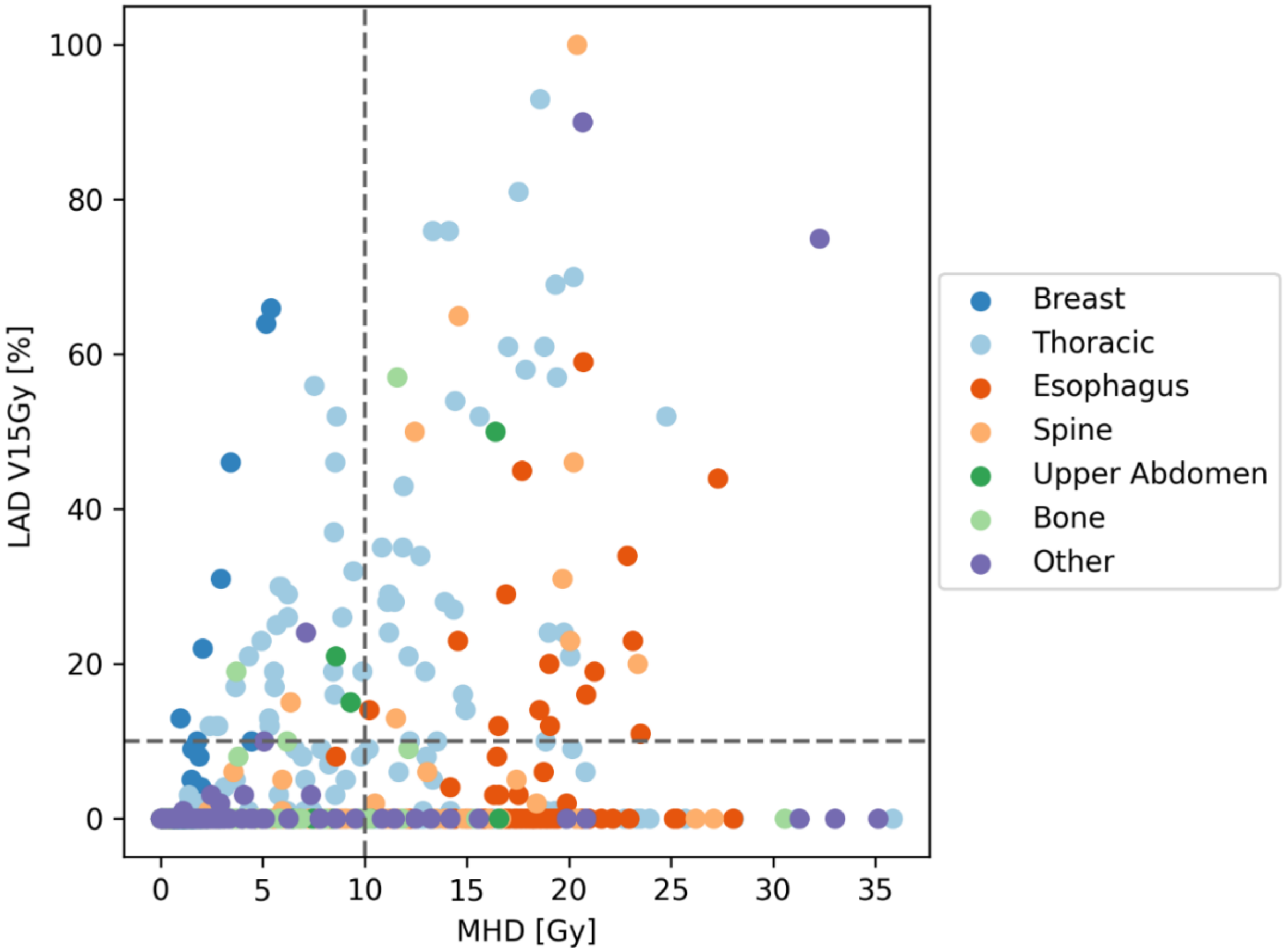

## TABLES

| Table 1. Baseline patient characteristics in study cohorts. | | | | | | | | |
|---|---|---|---|---|---|---|---|---|
| | **BWH/DFCI Training Set (n=560)** | **BWH/DFCI Internal Validation Set (n=69)** | **BWH/DFCI Internal Test Set (n=71)** | **CSMC External Validation, NSCLC (n=102)** | **CSMC External Validation, Breast Cancer (n=181)** | **Real-World Retrospective Lung Cohort (n=3254)** | **Prospective Clinical Deployment (n=1950)** | **Prospective Clinical Deployment Lung Cohort (n=475)** |
| **Median (Age, IQR)** | 65 (57-73) | 64 (55-73) | 66 (58-75) | 71 (64, 77) | 63 (53, 72) | 70 (62-77) | 65 (56-74) | 73 (64-79) |
| **Female sex** | 48.9% (274) | 47.8% (33) | 52.1% (37) | 54.9% (56) | 100% (181) | 56.6% (1842) | 73.2% (1427) | 55.8% (265) |
| **ECOG PS** | | | | | | | | |
| **0-1** | 87.5% (490) | 87.0% (60) | 91.5% (65) | 90.2% (90) | 98.9% (179) | - | - | - |
| **2** | 9.6% (54) | 13.0% (9) | 5.6% (4) | 6.9% (7) | 1.1% (2) | - | - | - |
| **>3** | 2.9% (16) | 0% (0) | 2.8% (2) | 2.9% (3) | - | - | - | - |
| **Race** | | | | | | | | |
| **White** | 91% (510) | 85.5% (59) | 88.7% (63) | - | - | 90% (2930) | 84.1% (1640) | 89.9% (427) |
| **Black** | 4.3% (24) | 8.7% (6) | 7.0% (5) | - | - | 3.5% (115) | 5.2% (101) | 3.2% (15) |
| **Other** | 7.5% (42) | 5.8% (4) | 4.2% (3) | - | - | 6.4% (209) | 10.7% (209) | 6.9% (33) |
| **Smoking** | | | | | | | | |
| **Current** | 40% (224) | 40.6% (28) | 38% (27) | 10.8% (11) | 3.3% (6) | 11.8% (384) | 5.8% (114) | 11.8% (56) |
| **Former** | 52.1% (292) | 50.7% (35) | 53.5% (38) | 65.7% (67) | 33.3% (60) | 63.7% (2074) | 33.4% (651) | 53.7% (255) |
| **Never** | 7.9% (44) | 8.7% (6) | 8.5% (6) | 23.5% (24) | 63.3% (114) | 13.2% (431) | 39.8% (777) | 18.1% (86) |
| **HTN** | 52.3% (293) | 52.2% (36) | 46.5% (33) | 64.7% (66) | 42.5% (77) | 50.8% (1652) | 36.7% (716) | 54.7% (260) |
| **DM** | 14.6% (82) | 11.6% (8) | 9.9% (7) | 29.4% (30) | 13.8% (25) | 15.2% (494) | 13.3% (260) | 18.7% (89) |
| **On Statin** | 42.7% (239) | 33.3% (23) | 39.4% (28) | 49.0% (50) | 35.0% (63) | 47.5% (1547) | 30.7% (598) | 49.9% (237) |
| **On Anti-HTN** | 42.3% (237) | 40.6% (28) | 36.6% (26) | 57.8% (59) | 43.9% (79) | 66% (2147) | 44.7% (872) | 65.3% (310) |
| **Any CHD** | 28.7% (161) | 29.0% (20) | 29.6% (21) | 31.4% (32) | 8.3% (15) | 26.2% (851) | 17.0% (332) | 30.9% (147) |
| **CAD** | | | | 31.4% (32) | 5.0% (9) | 5.1% (165) | 3.7% (73) | 8% (38) |
| **Stroke** | 2.3% (13) | 39.1% (27) | 0% (0) | 6.9% (7) | 0.6% (1) | 7.8% (255) | 2.9% (56) | 7.4% (35) |
| **Cancer Type** | | | | | | | | |
| **Lung** | 100% (560) | 100% (69) | 100% (71) | 100% (102) | - | 100% (3254) | 24.4% (475) | 100% (475) |
| **Breast** | - | - | - | - | 100% (181) | - | 49.8% (972) | - |
| **Esophagus** | - | - | - | - | - | - | 4.1% (80) | - |
| **Spine** | - | - | - | - | - | - | 7.6% (148) | - |
| **Upper abddomen** | - | - | - | - | - | - | 3.3% (65) | - |
| **Other** | - | - | - | - | - | - | 10.8% (210) | - |
| **RT Technique** | | | | | | | | |
| **3D-CRT** | 76.1% (426) | 79.7% (55) | 80.3% (57) | 19.6% (20) | 96.7% (175) | 27.6% (898) | 57.1% (1114) | 7.6% (36) |
| **IMRT** | 23.9% (134) | 20.3% (14) | 19.7% (14) | 80.4% (82) | 3.3% (6) | 32.9% (1072) | 21.9% (427) | 45.7% (217) |
| **SBRT** | - | - | - | - | - | 39.5% (1284) | 21.0% (409) | 46.7% (222) |
| **Median RT dose (IQR)** | 66 (56-66) | 63 (54-66) | 66 (60-66) | 60 (56-60) | 50 (43-50) | 54 (45-60) | 49 (39-55) | 55 (50-60) |
| **Median RT fractions (IQR)** | 33 (28-33) | 32 (27-33) | 33 (30-33) | 30 (30-33) | 25 (16-28) | 10 (5-27) | 16 (5-22.75) | 8 (5-25) |
| **Year of RT** | | | | | | | | |
| **≤2005** | 14.1% (79) | 13.0% (9) | 16.9% (12) | 1.0% (1) | 0 | 0 | 0 | 0 |
| **2006-2010** | 48.2% (270) | 50.7% (35) | 54.9% (39) | 11.8% (12) | 3.9% (7) | 0 | 0 | 0 |
| **2011-2015** | 37.7% (211) | 36.2% (25) | 28.2% (20) | 30.4% (31) | 47.5% (86) | 15.9% (517) | 0 | 0 |
| **2016-2020** | 0 | 0 | 0 | 52.0% (53) | 48.6% (88) | 58.1% (1889) | 0 | 0 |
| **2021-present** | 0 | 0 | 0 | 4.9% (5) | 0 | 26.1% (848) | 100% (1950) | 100% (475) |
| **DVH Data, median (IQR)** | | | | | | | | |
| **Mean Heart (Gy)** | 12.1 (5.8-19.1) | 11.2 (6.5-16.5) | 15 (6.2-19.2) | 11.4 (6.3-19.5) | 1.2 (0.7-2.2) | 6.1 (1.4-11.7) | 0.7 (0.2-2.0) | 4.6 (0.8-10.1) |
| **LAD V15 Gy (%)** | 12.9 (0-42.6) | 17.4 (0.5-37.4) | 15.3 (0-43.8) | 25.0 (0-44.0) | 0 (0-4.0) | 0 (0-22) | 0 (0-0) | 0 (0-0) |
| **LCX V15 Gy (%)** | 5.3 (1.8-14.5) | 5.3 (2.4-13.1) | 6 (2-18.4) | 14.5 (0-51.0) | 0 (0-0) | 0 (0-24) | 0 (0-0) | 0 (0-0) |
| **Mean Coronary (Gy)** | 7.4 (2.5-15.9) | 7.2 (3.2-15.2) | ? | ? | ? | 3.9 (0.8-10) | 0.7 (0.2-2.1) | 3.1 (0.6-8) |
| **High risk cardiac dose*** | 29.5 % (165) | 29% (20) | 35.2% (25) | ? | ? | 32% (1086) | 16.3% (317) | 34.1% (162) |
| **CV Risk Estimates** | | | | | | | | |
| **ChyLL Score, median (IQR)** | 5.7 (2.4-6.8) | 5.8 (3.4-6.8) | 5.9 (2.6-7.1) | 6.3 (range: 0-8.1) | - | 4.8 (1.3-6.8) | 0 (0-2.2) | 1.3 (0-6.8) |
| **CHyLL Score >5** | 61.4% (344) | 62.3% (43) | 67.6% (48) | 68.6% (70) | - | 49.2% (1602) | 22.6% (441) | 41.5% (197) |
| **Framing ham Group** | | | | | | | | |
| **Low** | 15.4% (86) | 27.5% (19) | 18.3% (13) | - | - | 7.5% (245) | 22.5% (439) | 5.9% (28) |
| **Inte rmediate** | 16.1% (90) | 11.6% (8) | 14.1% (10) | - | - | 11.7% (380) | 14% (273) | 11.8% (56) |
| **Hig h** | 23% (129) | 17.4% (12) | 15.5% (11) | - | - | 26.1% (850) | 17.9% (350) | 26.3% (125) |

***High risk exposure defined as mean heart dose** (MHD) $\geq$10 Gy, mean coronary dose $\geq$10 Gy, LAD V15 $\geq$10%, or LCX V15 $\geq$10%

Abbreviations: BWH=Brigham and Women's Hospital; CAD=coronary artery disease; CHD=coronary heart disease; CHyLL=Cardiac disease, Hypertension, and Logarithmic left anterior descending coronary artery dose risk score; CSMC=Cedars-Sinai Medical Center; DFCI=Dana-Farber Cancer Institute; ECOG PS=Eastern Cooperative Oncology Group performance status; Gy=gray; HTN=hypertension; IMRT=intensity-modulated radiation therapy; IQR=interquartile range; LAD=left anterior descending coronary artery; LCX=left circumflex coronary artery; RT=radiotherapy; SBRT=stereotactic body radiotherapy.

**Table 2**. Internal and external validation of AI-generated versus manual left anterior descending (LAD) coronary artery V15 Gy in BWH internal (n=70) and CSMC external validation (n=283) cohorts for clinical outcome prediction (please see Supplemental Tables for further details of prediction models).

| Fine and Gray regression to predict major adverse cardiac events (MACE). | | | | | |
|---|---|---|---|---|---|
| Institution / Dataset | Variable | Univariable | | Multivariable[#] | |
| | | HR (95% CI) | p | SHR (95% CI) | p |
| Internal (BWH) | Manual LAD V15 Gy per % | 1.02 (1.00-1.03) | .038 | **1.05 (1.03-1.07)** | **<.001** |
| Internal (BWH) | AI-Generated LAD V15 Gy per % | 1.02 (1.00-1.03) | .035 | **1.03 (1.01-1.05)** | **.014** |
| External (CSMC)* | AI-Generated LAD V15 Gy per % | 1.01 (0.99-1.02) | .040 | **1.02 (1.00-1.03)** | **.045** |
| Cox regression to predict all-cause mortality. | | | | | |
| Institution / Dataset | Variable | Univariable | | Multivariable* | |
| | | HR (95% CI) | p | SHR (95% CI) | p |
| Internal (BWH) | Manual LAD V15 Gy per % | 1.01 (0.99-1.02) | .26 | **1.03 (1.01-1.05)** | **.006** |
| Internal (BWH) | AI-Predicted LAD V15 Gy per % | 1.00 (0.99-1.01) | .59 | **1.02 (1.00-1.03)** | **.046** |
| External (CSMC) | AI-Predicted LAD V15 Gy per % | 1.02 (1.01-1.02) | <.001 | **1.02 (1.01-1.03)** | **.001** |

[#]MACE multivariable model includes age, sex, HTN, CHD history, and interaction term of CHD*LADV15Gy. *MACE multivariable model limited to CHD history, LAD V15 Gy, and interaction term of CHD*LADV15Gy due to limited events (n=26). Abbreviations: BWH, Brigham and Women's Hospital; CSMC, Cedars-Sinai Medical Center; SHR, sub-distribution hazard ratio.

Abbreviations: AI=artificial intelligence; BWH=Brigham and Women's Hospital; CI=confidence interval; CSMC=Cedars-Sinai Medical Center; Gy=gray; HR=hazard ratio; LAD=left anterior descending coronary artery; MACE=major adverse cardiac events; SHR=subdistribution hazard ratio; V15Gy=volume receiving ≥15 Gy.

## SUPPLEMENTAL METHODS

### *1. Cardiac Sub-Structure Deep Learning Auto-Segmentation Algorithm Training*

We developed a deep learning system for the automatic segmentation of cardiac substructures on planning CT scans, using expert annotations from thoracic radiation oncology. The goal of this tool is to enable detailed dosimetric assessment of cardiac exposure in lung radiotherapy, facilitating both research and clinical quality assurance in thoracic radiation planning.

The training data consisted of contrast-enhanced and non-contrast CT scans from lung cancer patients treated at Dana-Farber Cancer Institute and Brigham and Women's Hospital (DFCI-BWH). Expert delineations of cardiac substructures—the left and right atria, ventricles, ascending aorta, and coronary arteries—were performed by experienced radiation oncology residents and approved by senior radiation oncologists, using atlas definitions as part of an ongoing cardiac toxicity study that identified an association between coronary dose and risk of major adverse cardiovascular events[7].

In total, the model was trained and tuned using 630 annotated CT scans, with 70 additional scans reserved for independent evaluation. To ensure generalizability, the training cohort included diverse imaging data in terms of scanner model (Siemens and GE) and acquisition protocol with and without contrast use.

The system employs a two-stage 3D U-Net-based architecture. The first stage localizes the heart on downsampled CT scans, producing a coarse segmentation used to define a bounding box. The second stage then segments the cardiac substructures in high resolution within the cropped region.

The localization network operated on downsampled volumes of [e.g., 112×112×112 voxels, 3mm³ resolution], using data augmentation including translations (±10 voxels) and 3D rotations (±4°). The network was trained for 1,200 epochs with a 70/30 train-validation split.

The segmentation network received cropped volumes of [e.g., 384×384×80 voxels] around the heart and was down -sampled to [128×128×80 voxels] for memory efficiency. Training included extensive augmentation (±35° rotation, ±20 voxel translation), and the final model was trained over 1,000 epochs on the full training cohort. Outputs were up-sampled to original resolution to preserve anatomical detail.

Model development and training were conducted on a Linux-based workstation using TensorFlow 2.8.0 with GPU acceleration (CUDA v11.2). Two NVIDIA A6000 GPUs with 48GB RAM of memory were used for training the model.

We tested generalization by applying the model to an independent set of thoracic RT planning scans, including both contrast and non-contrast studies not seen during training. The system demonstrated robust segmentation performance across scanner types and acquisition parameters.

**2. Cardiac Risk Stratification Informatics Approach**

a. The Research Patient Data Registry (RPDR) repository at Mass General Brigham HealthCare was queried to gather various relevant cardiac risk factors. ICD9 and ICD10 codes were used to find past medical history used in calculating cardiac risk scores such as coronary heart disease (CHD) and equivalent diagnoses. The same database was sourced for physical exam variables, including blood pressure and BMI, and for each variable, the closest datapoint prior to the radiation start date was taken.

This process was similar to the query for smoking status, but a word search was also employed to filter through social history notes for a minority of the patients. All statin except nystatin and a list of common anti-hypertension medications approved by a cardiologist were extracted using the RPDR database.

b. Each patient's HLLCHyLL score was calculated with the below formula and a CHyLL score greater than or equal to 5 was labelled as high and less than 5 as low risk. The CHyLL score is a composite of: CHD includingincluding equivalent diagnoses, any history of pre-RT hypertension diagnosis (HTN), and LADV15 values (generated using AI auto-segmentation and the treated plan).[37]

CHyLL score calculation:

$$5.51CHD + 1.28HTN + 1.48\ln(LADV15 + 1) - 1x.36CHD \times \ln(LADV15 + 1)$$

c. Framingham 10 year risk score was calculated with below formula, and a score of less than 0.1 was labelled as low, a score of between 0.1 and 0.2 was labelled as intermediate and a risk score of above 0.2 was labelled as high.[38]

Framingham risk score:

$$\hat{p} = 1 - S_0(t)^{e^{\sum_{i=1}^{p} \beta_i X_i - \sum_{i=1}^{p} \beta_i \bar{X}_i}}$$

**SUPPLEMENTAL RESULTS**

***ROCardS Algorithm Performance : Processing and Segmentation Time***

The average processing and segmentation times for cardiac sub-structures was 51.5 ± 9.6 seconds and 3.0 ± 1.5 seconds, respectively. Processing time included sampling of the CT volume to conform to the necessary input for the deep learning network and final post-processing and sampling back to the original resolutions of the CT.

***External Clinical Validation: Cedars-Sinai Medical Center Dataset:***

The external validation cohort included 283 patients with lung cancer (locally advanced NSCLC; n=102) or breast cancer (n=181) treated with thoracic RT between 2005 to 2020 at Cedars-Sinai Medical Center.

The median age 67 years (IQR, 57-74), 83.8% female (n=237), 50.5% with hypertension (n=143), 16.6% with CHD (n=47), the mean LAD V15 Gy (%) was 16.7% (SD 24.5%).
Among patients with NSCLC, the median follow up was 23.5 months (IQR, 15.6-53.2), 54.9% were female (n=56), 64.7% had HTN (n=66), 31.4% had CHD (n=32), and predicted mean LAD V15 Gy was 26.6% (SD 26.1%).

Among patients with locally advanced NSCLC (n=102), the median follow up was 23.5 months (IQR, 15.6-53.2), median age 71 years (IQR, 64-77), 54.9% were female (n=56), 64.7% had hypertension (HTN; n=66), 31.4% had coronary heart disease (CHD; n=32), and AI-generated mean LAD V15 Gy was 26.6%±26.1%. Among patients with non-metastatic breast cancer (n=181), the median follow up was 69.9 months (IQR, 53.4-86.0 months), median age 63 years (IQR, 53-72), 100% were female, 42.5% had HTN (n=77), 8.3% had CHD (n=15), and the AI-predicted mean LAD V15 Gy was 9.4±18.9%.

There were 26 major adverse cardiovascular events (MACE) (n=11 lung, n=15 breast), the median time to first MACE was 56.1 months overall (IQR, 27.1-79.8) and 22.1 months in lung (IQR, 15.4-52.0) and 67.8 months for breast (IQR, 52.7-85.3). The 2-year cumulative incidence of MACE was 5.1% (95% CI 3.0-8.0%), including 9.8% (95% CI, 5.2-16.2%) for lung and 2.2% (95% CI, 0.7%-5.2%). Accounting for age, pre-existing HTN or CHD, there was an increased risk of MACE with predicted LAD V15 Gy ≥10% vs. <10% (subdistribution hazard ratio [SHR] 4.03, 95% CI 1.22-13.31; p=.022).

**SUPPLEMENTAL FIGURES**

**Supplemental Figure 1**. Example email automated heart quality assurance (QA) report issued by ambient functionality built into the DL-ROCardS. The results shown are for three anonymized patients (two thoracic RT and one non-thoracic) and include information about the patient, RT plan, prescription dose, comparison between AI prediction output and manually segmented heart contours (Dice), and dose metrics including mean heart dose, mean coronary dose, and V15 for LAD, LCX, and LV.

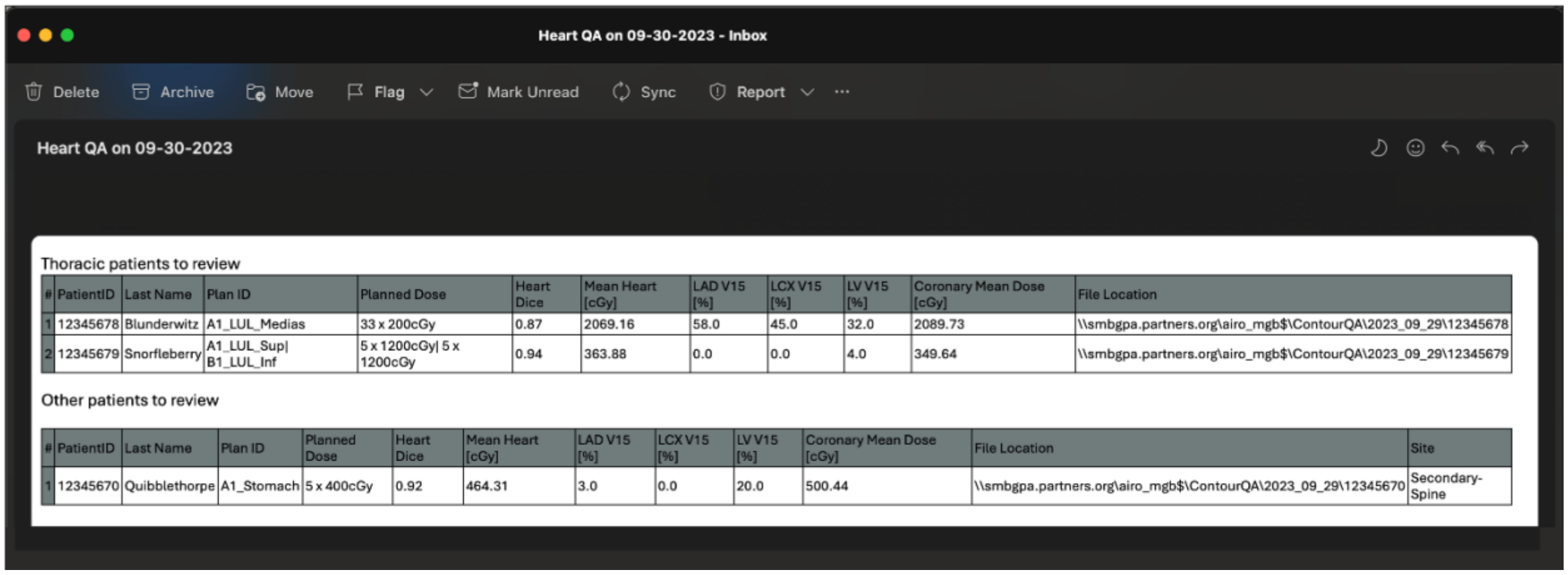

Thoracic patients to review

<table>
<tr><th>#</th><th>PatientID</th><th>Last Name</th><th>Plan ID</th><th>Planned Dose</th><th>Heart Dice</th><th>Mean Heart [cGy]</th><th>LAD V15 [%]</th><th>LCX V15 [%]</th><th>LV V15 [%]</th><th>Coronary Mean Dose [cGy]</th><th>File Location</th></tr>
<tr><td>1</td><td>12345678</td><td>Blunderwitz</td><td>A1_LUL_Medias</td><td>33 x 200cGy</td><td>0.87</td><td>2069.16</td><td>58.0</td><td>45.0</td><td>32.0</td><td>2089.73</td><td>\\smbgpa.partners.org\airo_mgb$\ContourQA\2023_09_29\12345678</td></tr>
<tr><td>2</td><td>12345679</td><td>Snorfleberry</td><td>A1_LUL_Sup|B1_LUL_Inf</td><td>5 x 1200cGy| 5 x 1200cGy</td><td>0.94</td><td>363.88</td><td>0.0</td><td>0.0</td><td>4.0</td><td>349.64</td><td>\\smbgpa.partners.org\airo_mgb$\ContourQA\2023_09_29\12345679</td></tr>
</table>

Other patients to review

| # | PatientID | Last Name | Plan ID | Planned Dose | Heart Dice | Mean Heart [cGy] | LAD V15 [%] | LCX V15 [%] | LV V15 [%] | Coronary Mean Dose [cGy] | File Location | Site |
|---|---|---|---|---|---|---|---|---|---|---|---|---|
| 1 | 12345670 | Quibblethorpe | A1_Stomach | 5 x 400cGy | 0.92 | 464.31 | 3.0 | 0.0 | 20.0 | 500.44 | \\smbgpa.partners.org\airo_mgb$\ContourQA\2023_09_29\12345670 | Secondary-Spine |

**Supplemental Figure 2**. Representative “best” (top row) and “worst” (bottom row) case examples of cardiac sub-structure auto-segmentations generated by DL-ROCardS on the internal validation dataset, including left ventricle (LV), right ventricle (RV), left atrium (LA), right atrium (RA), left anterior descending (LAD), left circumflex (LCX), right (RCA), and posterior descending (PDA) coronary arteries. The “best” case had a mean Dice Coefficient of 0.91 across all structures while the “worse” case had a mean Dice of 0.81. The first column demonstrates a 3D rendering of the 4 cardiac chambers, the second column demonstrates a 3D rendering of the coronaries, and the third column demonstrates a representative axial CT image from the two patients with AI-generated cardiac substructure segmentations overlaid. In the “worst” case example, discontinuities in the segmentations of smaller structures (e.g. left anterior descending [LAD[ coronary) can be seen.

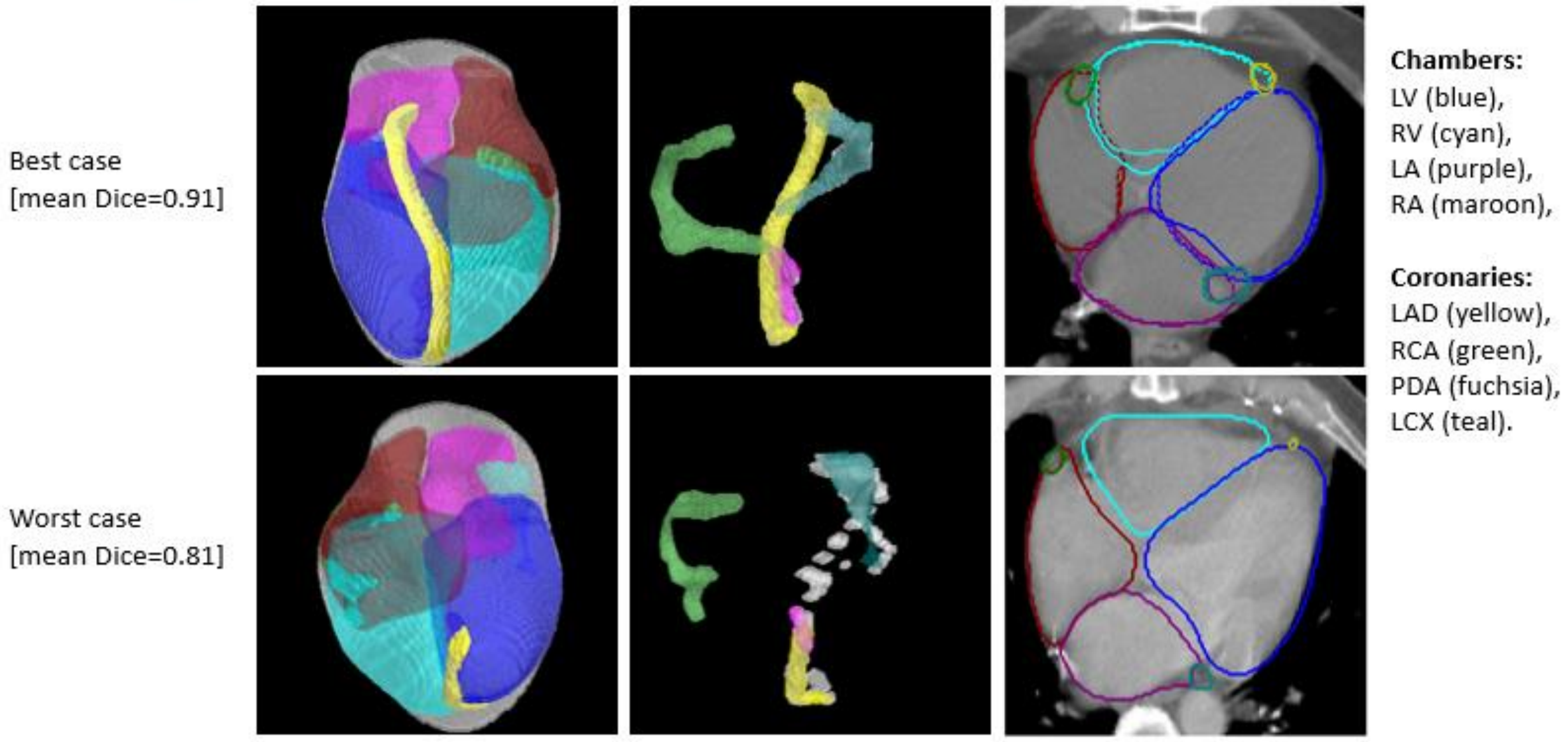

**Supplemental Figure 3.** Kaplan-Meier estimates of all-cause mortality in CSMC external validation cohort (n=283) stratified by AI-predicted LAD V15 Gy ≥10% vs <10% (log-rank p<.001).

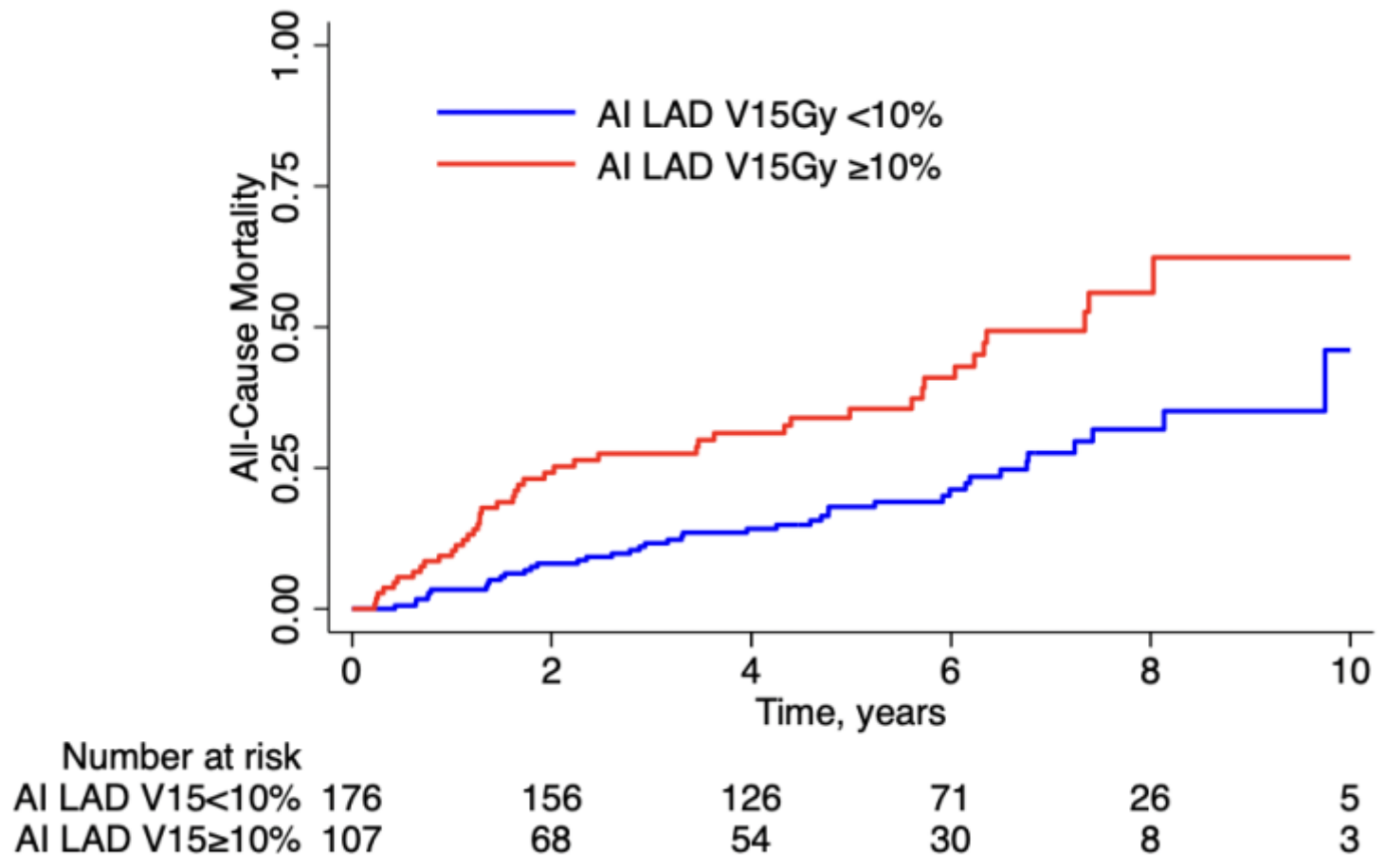

**Supplemental Figure 4**. Violin plots of radiation dose exposures to the left anterior descending (LAD) and left circumflex (LCX) coronary arteries (V15 Gy), mean heart (MHD), and mean total coronary in the prospective ambient surveillance cohorts (n=1,950).

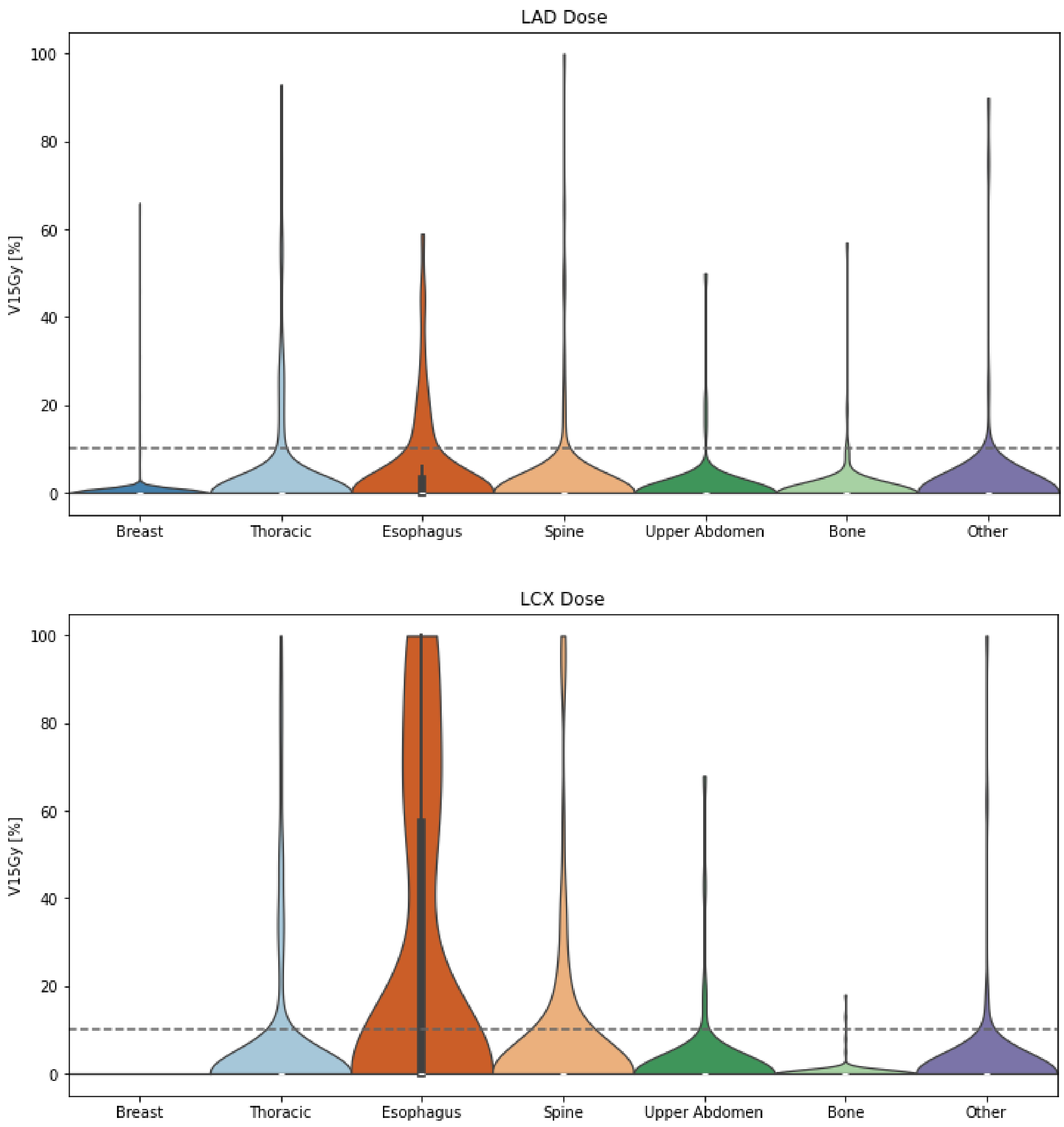

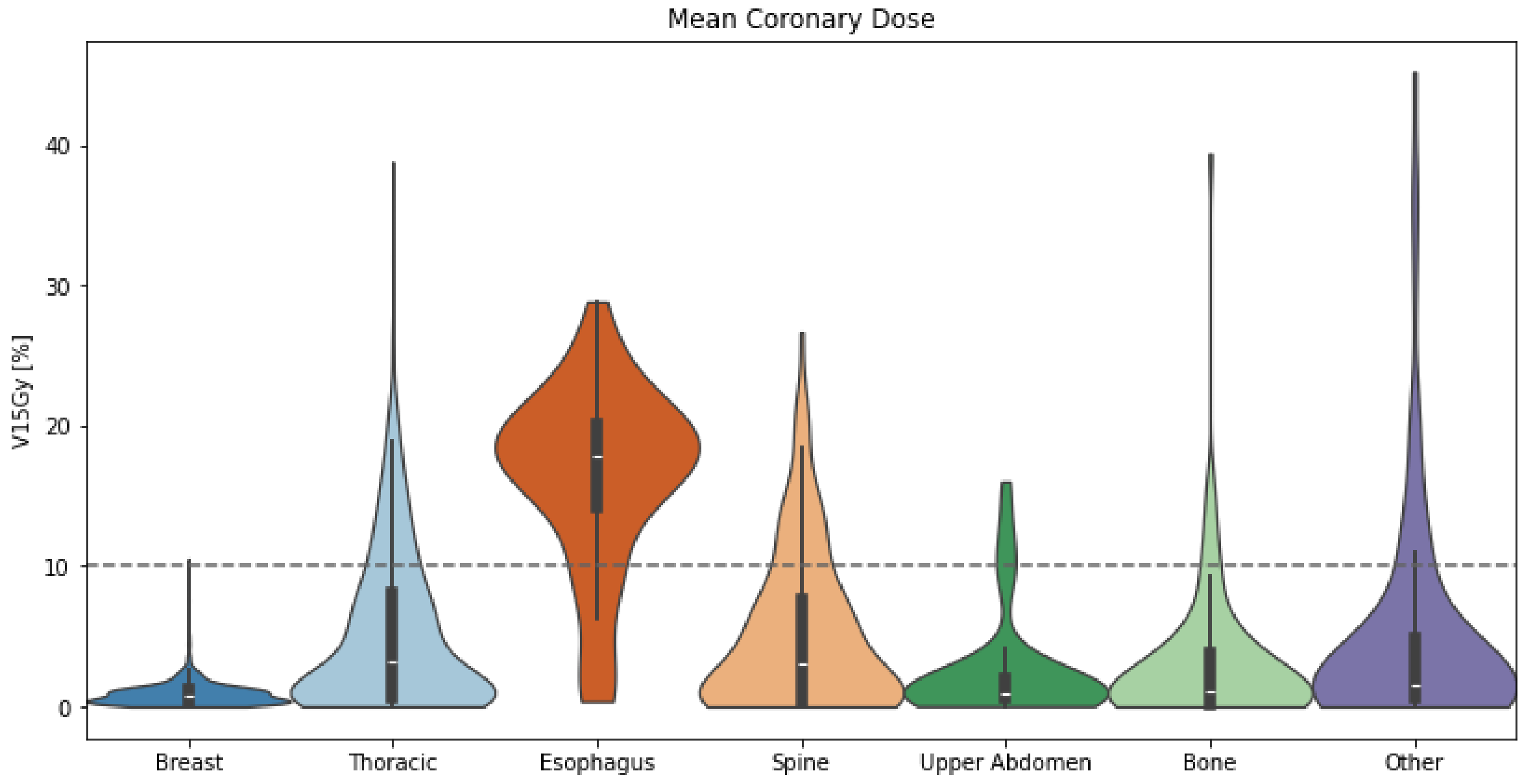

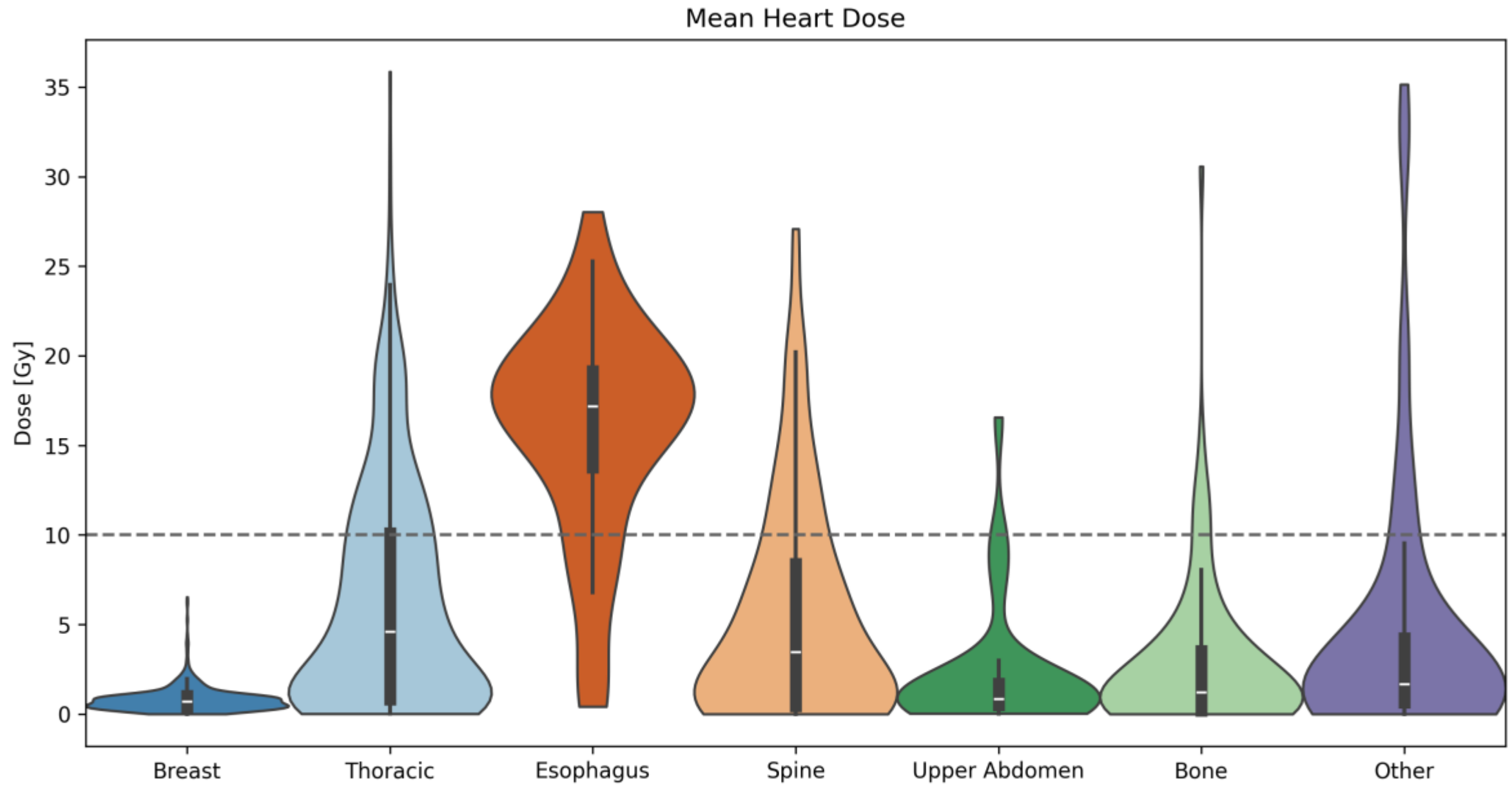

## SUPPLEMENTAL TABLES

**Supplemental Table 1.** Median and interquartile range (parentheses) for volumetric and dosimetric comparisons between manual and AI-generated cardiac sub-structures in the BWH/DFCI internal validation dataset.

| **Internal Validation Dataset (n=70)** | | | | | | | | | | |
|---|---|---|---|---|---|---|---|---|---|---|
| **Name** | Heart | LA | RA | LV | RV | LAD | LCX | LM | RCA | PDA |
| Dice | 0.95 (0.01) | 0.86 (0.04) | 0.85 (0.05) | 0.90 (0.03) | 0.84 (0.06) | 0.65 (0.09) | 0.55 (0.18) | 0.40 (0.29) | 0.56 (0.14) | 0.47 (0.22) |
| ASSD (mm) | 1.20 (0.53) | 1.60 (0.53) | 1.07 (0.65) | 1.60 (0.69) | 2.10 (0.94) | 1.20 (0.60) | 1.90 (1.20) | 1.80 (1.90) | 1.90 (0.95) | 1.80 (1.90) |
| Mean Dose (Gy) | -0.03 (0.80) | 0.06 (0.14) | -0.01 (0.60) | -0.04 (0.30) | -0.16 (1.28) | -0.01 (0.14) | -0.19 (1.26) | -0.00 (2.26) | 0.02 (0.92) | -0.00 (0.10) |
| V15Gy | -0.01 (0.02) | 0.00 (0.04) | 0.00 (0.03) | -0.01 (0.02) | -0.03 (0.04) | -0.01 (0.05) | -0.01 (0.10) | 0.00 (0.64) | 0.00 0.07 | -0.05 (0.10) |
| Global DVH metric (IQR)* | 0.60 (0.83) | 0.47 (0.89) | 0.18 (0.78) | 0.16 (0.96) | 0.06 (0.51) | 0.09 (0.66) | 0.16 (0.93) | 0.03 (0.87) | 0.22 (0.89) | 1.00 (0.13) |

* Kolmogorov-Smirnov median p-value (P>0.05 suggests no significant deviation of distribution in the radiation dose-volume histogram for AI versus manual segmentation for a given organ)

Abbreviations: ASSD=average symmetric surface distance; Dice=Dice similarity coefficient; DVH=dose–volume histogram; Gy=gray; IQR=interquartile range; LA=left atrium; LAD=left anterior descending coronary artery; LCX=left circumflex coronary artery; LM=left main coronary artery; LV=left ventricle; mm=millimeters; PDA=posterior descending artery; RA=right atrium; RCA=right coronary artery; RV=right ventricle; V15Gy=volume receiving ≥15 Gy.

| **Supplemental Table 2**. Univariable and multivariable MACE analysis in BWH Validation Cohort (n=70) using AI-generated LAD V15 Gy.<br>Covariable | Univariable | | Multivariable | |
|---|---|---|---|---|
| | HR (95% CI) | p | SHR (95% CI) | p |
| Age | 1.07 (1.01-1.13) | .018 | 1.06 (0.98-1.14) | .13 |
| Sex (M vs F) | 0.63 (0.19-2.08) | .44 | | |
| Hypertension | 9.28 (1.16-73.91) | .035 | 4.34 (0.62-30.43) | .14 |
| CHD | 3.99 (1.22-13.06) | .022 | 3.59 (0.90-14.32) | .07 |
| AI-Predicted LAD V15 Gy (per %) | 1.02 (1.00-1.03) | .035 | **1.03 (1.01-1.05)** | **.014** |
| CHD*LADV15 | 0.98 (0.96-1.01) | .25 | 0.98 (0.96-1.01) | .20 |

Abbreviations: AI: artificial intelligence; CHD:coronary heart disease; LAD: left anterior descending coronary artery; MACE: major adverse cardiovascular event; SHR: subdistribution hazard ratio; V15: volume receiving 15 Gy

**Supplemental Table 3**. Univariable and multivariable MACE analysis in BWH Validation Cohort (n=70) using LAD V15 Gy calculated from manual LAD segmentation

| covariable | Univariable | | Multivariable | |
|---|---|---|---|---|
| | HR (95% CI) | p | SHR (95% CI) | p |
| Age | 1.07 (1.01-1.13) | .018 | 1.05 (0.97-1.13) | .23 |
| Sex (M vs F) | 0.63 (0.19-2.08) | .44 | | |
| Hypertension | 9.28 (1.16-73.91) | .035 | 7.17 (1.02-50.56) | .048 |
| CHD | 3.99 (1.22-13.06) | .022 | 6.46 (1.43-29.18) | .015 |
| Actual LAD V15 Gy (per %) | 1.02 (1.00-1.03) | .038 | **1.05 (1.03-1.07)** | **<.001** |
| CHD*LADV15 | 0.97 (0.94-1.00) | .08 | 0.96 (0.93-0.99) | .005 |

Abbreviations: AI: artificial intelligence; CHD:coronary heart disease; LAD: left anterior descending coronary artery; MACE: major adverse cardiovascular event; SHR: subdistribution hazard ratio; V15: volume receiving ≥15 Gy

**Supplemental Table 4**. Univariable and multivariable all-cause mortality (ACM) analysis in BWH validation cohort (n=70) using AI-generated LAD V15 Gy.

| Covariable | Univariable | | Multivariable | |
|---|---|---|---|---|
| | HR (95% CI) | p | aHR (95% CI) | p |
| Age | 1.02 (1.00-1.05) | .10 | 1.01 (0.98-1.04) | .56 |
| Sex (M vs F) | 0.93 (0.52-1.66) | .80 | 0.71 (0.37-1.36) | .30 |
| ECOG PS (0-1 vs ≥2) | 1.89 (0.80-4.48) | .15 | 2.21 (0.79-6.22) | .13 |
| Weight loss | 1.35 (0.64-2.83) | .43 | 1.59 (0.65-3.89) | .31 |
| Surgery | 0.29 (0.15-0.58) | <.001 | 0.23 (0.10-0.50) | <.001 |
| Prior MI | 1.66 (0.70-3.94) | .25 | 2.65 (0.75-9.31) | .13 |
| Prior CHF | 0.64 (0.20-2.08) | .46 | 0.24 (0.06-1.01) | .052 |
| Any CHD | 1.24 (0.69-2.25) | .47 | 1.80 (0.66-4.92) | .25 |
| RT after 2008 (vs. pre-2008) | 0.60 (0.34-1.08) | .09 | 0.65 (0.32-1.32) | .23 |
| AI-Predicted LAD V15 Gy per % | 1.00 (0.99-1.01) | .59 | **1.02 (1.00-1.03)** | **.046** |
| CHD*LADV15 | 0.99 (0.97-1.01) | .47 | 0.97 (0.94-1.00) | .031 |

Abbreviations: AI: artificial intelligence; CHD: coronary heart disease; CHF: congestive heart failure; LAD: left anterior descending coronary artery; MACE: major adverse cardiovascular event; MI: myocardial infarction; SHR: subdistribution hazard ratio; V15: volume receiving $\geq$ 15 Gy

**Supplemental Table 5**. Univariable and multivariable all-cause mortality (ACM) analysis in BWH validation cohort (n=70) using LAD V15 Gy calculated from manual LAD segmentation.

| Covariable | Univariable | | Multivariable | |
|---|---|---|---|---|
| | HR (95% CI) | p | aHR (95% CI) | p |
| Age | 1.02 (1.00-1.05) | .10 | 1.01 (0.98-1.04) | .63 |
| Sex (M vs F) | 0.93 (0.52-1.66) | .80 | 0.68 (0.35-1.31) | .25 |
| ECOG PS (0-1 vs ≥2) | 1.89 (0.80-4.48) | .15 | 1.96 (0.69-5.61) | .21 |
| Weight loss | 1.35 (0.64-2.83) | .43 | 1.55 (0.64-3.77) | .33 |
| Surgery | 0.29 (0.15-0.58) | <.001 | 0.22 (0.10-0.48) | <.001 |
| Prior MI | 1.66 (0.70-3.94) | .25 | 2.65 (0.75-9.42) | .13 |
| Prior CHF | 0.64 (0.20-2.08) | .46 | 0.24 (0.06-1.03) | .056 |
| Any CHD | 1.24 (0.69-2.25) | .47 | 2.08 (0.74-5.86) | .17 |
| RT after 2008 (vs. pre-2008) | 0.60 (0.34-1.08) | .09 | 0.66 (0.32-1.33) | .24 |
| Manual LAD V15 Gy per % | 1.01 (0.99-1.02) | .26 | **1.03 (1.01-1.05)** | **.006** |
| CHD*LADV15 | 0.98 (0.96-1.01) | .15 | 0.96 (0.93-0.99) | .007 |

Abbreviations: AI: artificial intelligence; CHD: coronary heart disease; CHF: congestive heart failure; LAD: left anterior descending coronary artery; MACE: major adverse cardiovascular event; MI: myocardial infarction; SHR: subdistribution hazard ratio; V15: volume receiving $\geq$ 15 Gy

**Supplemental Table 6.** Median and interquartile range (parentheses) for volumetric and dosimetric comparisons between manual and AI-generated cardiac sub-structures in the CSMC external dataset.

| **Name** | Heart | LV | LAD | LCX |
|---|---|---|---|---|
| **Dice** | 0.92 (0.06) | 0.87 (0.04) | 0.66 (0.15) | 0.47 (0.11) |
| ASSD (mm) | 2.2 (1.8) | 2.00 (1.10) | 1.3 (0.75) | 2.7 (1.4) |
| Mean Dose (Gy) | -0.27 (1.89) | 0.04 (0.25) | 0.33 (2.06) | -0.01 (0.60) |
| V15Gy | -0.01 (0.04) | 0.00 (0.03) | -0.03 (0.05) | -0.06 (0.08) |

Abbreviations: ASSD=average symmetric surface distance; Dice=Dice similarity coefficient; Gy=gray; LAD=left anterior descending coronary artery; LCX=left circumflex coronary artery; LV=left ventricle; mm=millimeters; V15Gy=volume receiving ≥15 Gy.

**Supplemental Table 7**: Univariable and multivariable MACE analysis in CSMC validation cohort (n=283) using AI-generated LAD V15 Gy.

| Covariable | Univariable | | Multivariable | |
|---|---|---|---|---|
| | HR (95% CI) | p | SHR (95% CI) | p |
| CHD | 6.17 (2.87-13.30) | <.001 | 11.50 (4.73-27.99) | <.001 |
| AI-Predicted LAD V15 (per %) | 1.01 (0.99-1.02) | .40 | **1.02 (1.00-1.03)** | **.045** |
| CHD*LAD V15Gy | | | 0.97 (0.94-0.99) | .014 |

Abbreviations: AI=artificial intelligence; CHD=coronary heart disease; CI=confidence interval; CSMC=Cedars-Sinai Medical Center; Gy=gray; HR=hazard ratio; LAD=left anterior descending coronary artery; MACE=major adverse cardiac events; SHR=subdistribution hazard ratio; V15Gy=volume receiving ≥15 Gy.

**Supplemental Table 8**. Univariable and multivariable all-cause mortality (ACM) analysis in CSMC validation cohort (n=283) using AI-Predicted LAD V15 Gy.

| Covariable | Univariable | | Multivariable | |
|---|---|---|---|---|
| | HR, 95% CI | p | AHR, 95% CI | p |
| Age | 1.03 (1.01-1.04) | .009 | 1.01 (0.99-1.03) | .42 |
| Sex (M vs F) | 0.36 (0.22-0.59) | <.001 | 0.49 (0.28-0.86) | .012 |
| ECOG PS (0-1 vs ≥2) | 1.74 (1.12-2.71) | .014 | 1.77 (1.12-2.79) | .014 |
| HTN | 1.50 (0.98-2.31) | .063 | 0.94 (0.57-1.56) | 0.81 |
| CHD | 2.22 (1.36-3.65) | .002 | 2.44 (1.20-4.97) | .014 |
| AI-predicted LAD V15 Gy ≥10% vs <10% | 1.02 (1.01-1.02) | <.001 | **1.02 (1.01-1.03)** | **.001** |
| CHD*LADV15 | - | - | .98 (0.96-0.99) | .043 |

Abbreviations: AHR=adjusted hazard ratio; AI=artificial intelligence; ACM=all-cause mortality; CHD=coronary heart disease; CI=confidence interval; CSMC=Cedars-Sinai Medical Center; ECOG PS=Eastern Cooperative Oncology Group performance status; F=female; Gy=gray; HR=hazard ratio; HTN=hypertension; LAD=left anterior descending coronary artery; M=male; V15Gy=volume receiving ≥15 Gy.

**Supplemental Table 9:** Stratification by High versus Low-Risk LAD Radiation Dose Exposure Calculated from AI-Generated Contours in the Lung Cancer Retrospective Cohort (n=3399) and Identifying Number of Patients with Coronaries Segmented and Constrained as Part of Clinical Care

| | Number of Patients with LAD V15 > 10% (% of total) | Number of Patients with LAD V15 $\leq$ 10% (% of total) | **Total** |
|---|---|---|---|
| Number of Patients with Any Coronaries Segmented by Clinicians as Part of Standard Clinical Care | 21 (0.6%) | 59 (1.7%) | 80 (2.3%) |
| Number of Patients with No Coronaries Segmented by Clinicians as Part of Standard Clinical Care | 1065 (31.3%) | 2253 (66.3%) | 2253 (97.6%) |
| Number of Patients with LAD Segmented and Constrained by Clinicians in the Radiation Oncology Prescription | 10 (0.3%) | 32 (0.9%) | 42 (1.2%) |
| **Total Patients** | 1086 (32.0%) | 2312 (68.0%) | 3398 |